\documentclass[preprint,amsmath,superscriptaddress,pre]{revtex4}  % Ramon modified
\usepackage[dvips]{graphicx}
\usepackage{epsfig}
\usepackage{color}
\usepackage{subfigure}
\usepackage{soul}

\begin{document} 

\title{
Tail of the distribution of fatalities in epidemics
%Scientific comment on 
%``Tail risk of contagious diseases''
}
\author{\'Alvaro Corral}
\affiliation{%
Centre de Recerca Matem\`atica,
Edifici C, Campus Bellaterra,
E-08193 Barcelona, Spain
}\affiliation{Departament de Matem\`atiques,
Facultat de Ci\`encies,
Universitat Aut\`onoma de Barcelona,
E-08193 Barcelona, Spain
}\affiliation{Barcelona Graduate School of Mathematics, 
Edifici C, Campus Bellaterra,
E-08193 Barcelona, Spain
}\affiliation{Complexity Science Hub Vienna,
Josefst\"adter Stra$\beta$e 39,
1080 Vienna,
Austria
}
\email{alvaro.corral@uab.es}
%\section*{Abstract}
\begin{abstract} 
%The size that an epidemic can reach,
The final size reached by an epidemic,
measured in terms of the total number of fatalities, 
is an extremely relevant quantity.
It has been recently claimed that 
%Cirillo and Taleb \cite{Cirillo_Taleb} study 
the size distribution of major epidemics in human history
%(including ongoing COVID-19)
%in terms of the number of fatalities.
%Using the figures from 72 epidemics (REPETIDO!!!), from the plague of Athens (429 BC)
%to the COVID-19 (2019-2020), they claim that the resulting 
%fatality distribution 
is ``strongly fat-tailed'', i.e., a power law asymptotically, 
which has important consequences for risk management.
%as the mean value of the fatality distribution becomes infinite.
%
%CONSEQUENCES??
%
%Power-law distributions are indeed an important paradigm in complex systems,
%representing the non-existence of characteristic scales.
%However, the empirical treatment of power-law data is a delicate issue
%\cite{White,Bauke, Clauset,corral_nuclear,Corral_Deluca,Hanel_power_laws,Corral_Gonzalez,Voitalov_krioukov},
%resulting in that the rigorous
%evidence in support of these distributions 
%has been rather limited \cite{Corral_Gonzalez}.
%
From the point of view of statistical physics and
complex-systems modeling this is not an unexpected outcome, 
nevertheless,
strong empirical evidence is also necessary to support such a claim. 
Reanalyzing previous data, %as Cirillo and Taleb, % \cite{Cirillo_Taleb}, 
we find that, although the fatality distribution may be compatible with a power-law tail,
%and in particular with a Pareto distribution, 
these results are not conclusive, 
and other distributions, not fat-tailed, could explain the data equally well.
As an example, 
simulation of a log-normally distributed random variable
provides synthetic data whose statistics are undistinguishable from the statistics of the
empirical data.
Theoretical reasons justifying a power-law tail
as well as limitations in the current available data
are also discussed.
%OJO A PARETO!!!
\end{abstract} 
\maketitle

\section{Introduction}

In complex systems, 
the statistical variability of
the size of a phenomenon carries fundamental information about its underlying dynamics
\cite{Bak_book,Mitz,Newman2004a,Sornette_critical_book,Thurner_book}.
Recently,
Cirillo and Taleb \cite{Cirillo_Taleb} have studied the size of major epidemics in human history, 
measured in number of fatalities.
Using the figures from 72 epidemics, from the plague of Athens (429 BC)
to the COVID-19 (still ongoing), they claim that the resulting 
fatality distribution is ``extremely fat-tailed'', i.e., a power law, asymptotically
(for very large values of the number of fatalities).
%
%CONSEQUENCES??
%
Power-law distributions are an important paradigm in complex-systems science,
representing the non-existence of characteristic scales
and the divergence of moments.
However, the empirical treatment of power-law data is a delicate issue
\cite{White,Bauke, Clauset,corral_nuclear,Corral_Deluca,Hanel_power_laws,Corral_Gonzalez,Voitalov_krioukov},
resulting in that the rigorous
evidence in support of these distributions 
has been rather limited \cite{Corral_Gonzalez}.

Reanalyzing the same epidemic data as Ref. \cite{Cirillo_Taleb}, 
we find that, although the data may be compatible with an asymptotic power-law,
%and in particular with a Pareto distribution, 
there is no 
%enough
evidence in favor of this behavior, 
%the results are not conclusive, 
and other distributions, not fat-tailed, could explain the tail equally well or even better.
In concrete, simulation of a tail coming from a truncated log-normal distribution 
provides data whose statistics are undistinguishable from the
epidemic empirical data.
Based on this log-normal tail we show how to provide a very rough estimate
of the final expected size of the COVID-19 pandemic (which would turn out to be infinite in the case of a power law).
%% calcular la media para comparar con los momentos fantasmas de taleb
We finally discuss under which physical circumstances one could expect a power-law tail 
for the size of epidemics, and the necessity of much better data. 
%OJO A PARETO!!!
The present paper significantly extends and puts in a wider context
the results of Ref. \cite{Corral_comment_cirillo_taleb}. 

\section{Data and completeness}

Most of the events analyzed in Ref. \cite{Cirillo_Taleb} come from the 
Wikipedia list 
%collected by Wikipedia
of the biggest known epidemics %(including pandemics) 
caused by infectious diseases \cite{Wikipedia_epidemics}.
The list was complemented with a few more events, 
obtained from other sources (mainly Ref. \cite{list_epidemics}). % mainly 24
Restricting to epidemics causing 1000 or more fatalities, 
this led to $N=72$ events, from 429~BC to 2020~AD (CE);
the resulting data is provided in Table 1 of Ref. \cite{Cirillo_Taleb},
in concrete, in the 4th numeric column (average estimate).
We will refer to this data as the original data set.

\begin{figure}[!ht]
\begin{center}
(a)
\includegraphics[width=7cm]{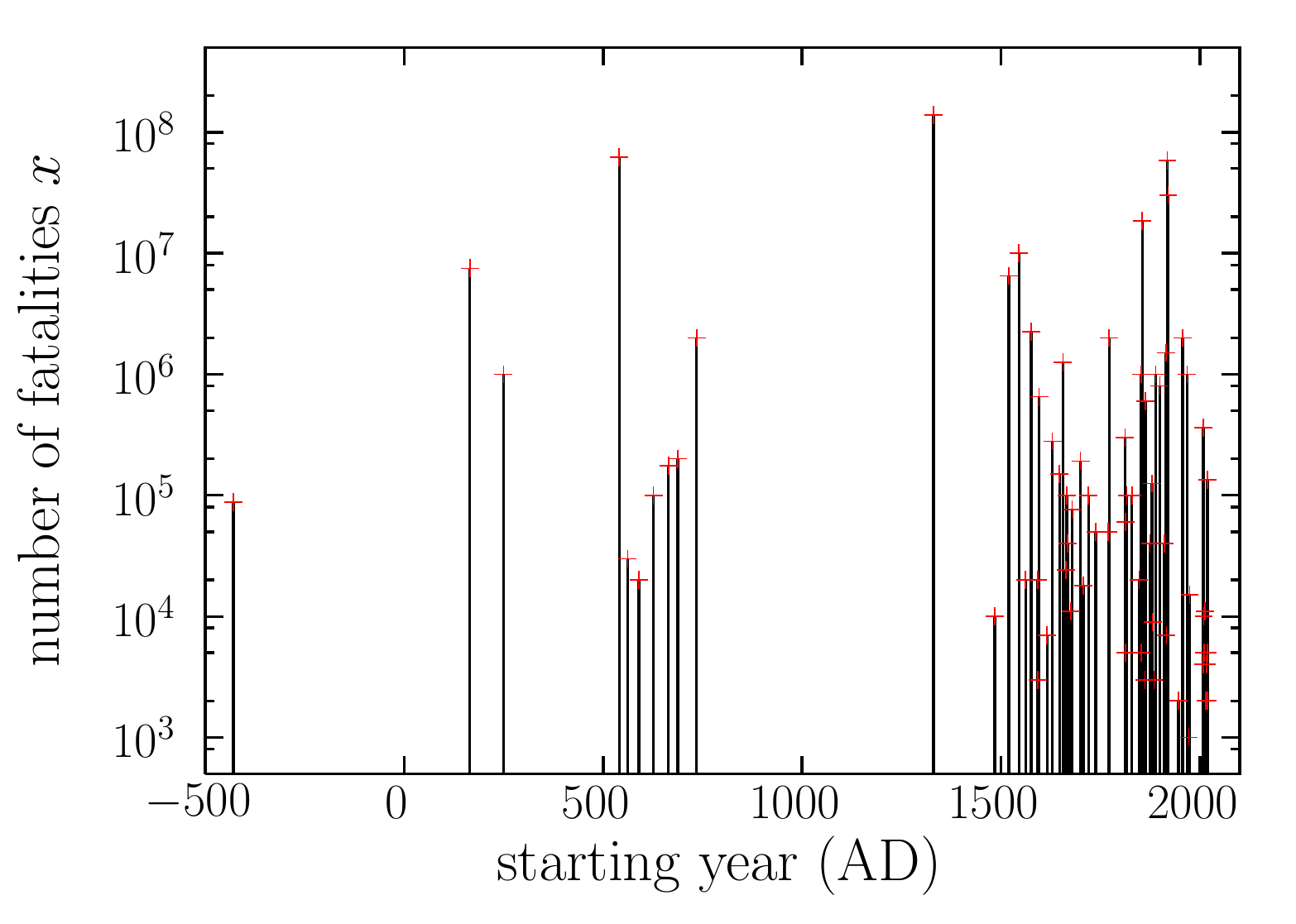}%{./Fig1a.pdf}
(b)
\includegraphics[width=7cm]{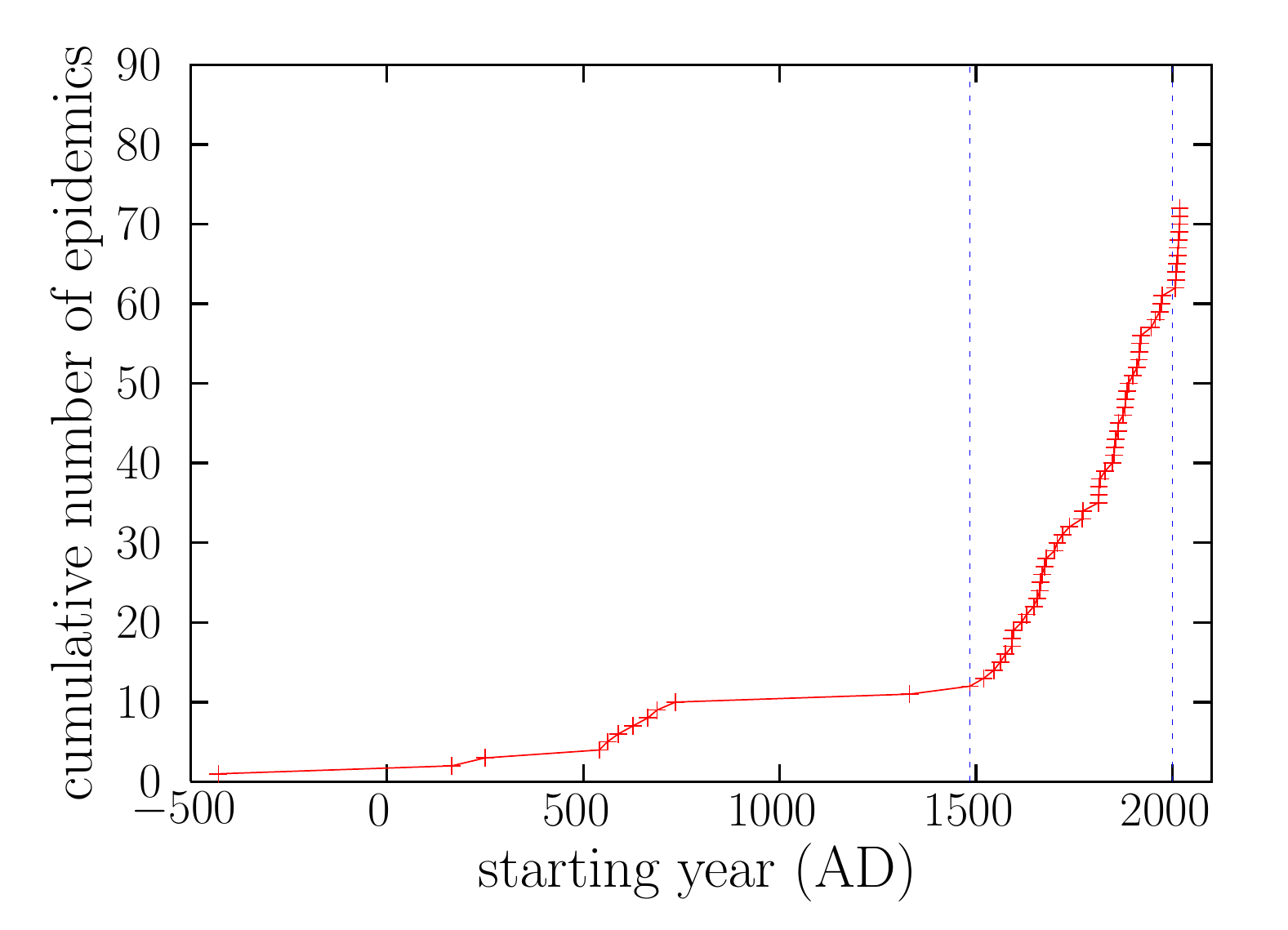}%{./Fig1b.pdf}
\end{center}
\caption{
Visualization of the original data set introduced in Ref. \cite{Cirillo_Taleb} and reanalyzed in the present paper.
(a) Number of fatalities of each epidemic versus year of occurrence
(beginning of the outbreak).
(b) Cumulative number of epidemics versus year of occurrence.
The slope yields the rate of occurrence.
The vertical bars mark the window selected for the restricted data set also analyzed in the present paper.
Both the inhomogeneity in time and the incompleteness in size become apparent.
}
\label{Fig1}
\end{figure}

We are not aware of any study of completeness of such data.
In order to address this important point,
we display the number of fatalities versus time of occurrence
(beginning of the outbreak) in Fig. \ref{Fig1}(a),
where it can be clearly seen that
the data looks certainly 
%incomplete and 
inhomogeneous.
For example, between 735~AD (Japanese smallpox epidemic)
and 1485~AD there is only one event, 
which happens to be the Black Death (largest event on record, 
with roughly 140,000,000 fatalities, starting in 1331).

%We also see that no event below 10,000 fatalities is recorded before 1485.

Figure \ref{Fig1}(b) confirms that the rate of occurrence of large epidemics
seems to be highly inhomogeneous, 
from roughly 3 per century between 500 and 750~AD % 7 en total
%(for a number of fatalities above 10,000)
to just 1 in seven centuries  (up to 1485)
and
to about 0.5 per year after 2000~AD. 
%(for fatalities above $1000$;
%this turns to 0.2 per year for fatalities $\ge$ 10,000).
We associate this high variability mainly to the large incompleteness of the data
and not to a wild variation of the occurrence of epidemics over the centuries.

Incompleteness can also be seen in the uneven distribution of the smallest events considered,
i.e., out of the 17 epidemics below 10,000 fatalities, 
only 2 are contained in the first half of the data (from 429 BC to 1813).
Including events below the completeness threshold can lead to important biases.
Although in Ref. \cite{Cirillo_Taleb} it is claimed that a Jackknife
resampling technique can account for incompleteness, 
it is obvious that resampling does not correct any bias present in the data
(one must not get confused by the fact that Jackknife corrects the bias of an estimator, 
which is clearly different from the bias arising from data incompleteness).
% https://es.wikipedia.org/wiki/Jackknife_(estad%C3%ADstica)

For this reason, in order to evaluate the impact of data incompleteness, 
in addition to analyze the raw data used in Ref. \cite{Cirillo_Taleb}, 
we also consider a restricted data set, from 1485 to 2000~AD,
which shows a higher homogeneity, with an average of 0.1 epidemics per year.
%(for fatalities $\ge$ 1000; 
%nevertheless, the number of epidemics with fatalities below 10,000
%is only ``non-negligible'' after 1800~AD). 
This restricted data set consists of $N=50$ epidemics.
Note that the largest event, the Black Death, is not contained in this data set;
thus, 
this provides a way to test the robustness of the results in front of the value of
the most extreme event.
This also eliminates ongoing epidemics (such as the COVID-19) from the analysis (except the HIV/AIDS pandemic). 

A third data set, considered in Ref. \cite{Cirillo_Taleb}, consist of rescaling 
the number of fatalities (of the original data set) by the whole world population 
at the time of the epidemic.
This represents a procedure to take into account that the world population is far from stationary
across history, and it is not the same to have 140 million of fatalities nowadays than in the Middle Ages (where the global population was around 400 million).

However, it is questionable why one should rescale by the world population
and not by the local population (i.e., the population of Greece for the plague of Athens),
and also why the scaling is linear with the total world population
(for instance, in a simple branching process the size does not scale linearly with the size
of the underlying tree, but with the square of the number of generations \cite{Corral_garcia_moloney_font}, 
and in general one has to take into account the so-called fractal dimension of the avalanches, see e.g. Ref. \cite{Corral_Paczuski}). %% Christensen_oslo}).
Nevertheless, we will also explore this simple rescaled data set, 
in order to test the robustness of the results. 

%%CITAR LA SECCION DEL FINITE SIZE SCALING!!!

\section{Distributions and tails}

The authors of Ref.
%Cirillo and Taleb 
\cite{Cirillo_Taleb} identify ``fat-tailed'' distributions 
with regularly varying distributions \cite{Voitalov_krioukov}, 
defined by a complementary cumulative distribution function
(or survival function \cite{footnote_survival}, probability of being above $x$) 
given by 
$S_{fat}(x) = \ell(x)/x^{\alpha}$, 
with $\alpha$ the exponent 
(of $S_{fat}(x)$, and $\xi=1/\alpha$ the tail index, measuring the ``fatness'' of the tail)
and $\ell(x)$ an unspecified slowly varying function
(for example, a function that tends to a constant when $x \rightarrow \infty$, 
but not only).
Roughly speaking, a ``fat-tailed'' distribution becomes a power law asymptotically.

From the terminological point of view, the name ``fat tail''
becomes unfortunate, as it can be confused with the terms 
``long tail'' and ``heavy tail'', which denote different classes of probability distributions.
It is convenient to remember that
fat-tailed distributions are subexponential distributions, 
which are long-tailed in their turn, which are heavy-tailed, 
but not the opposite \cite{Voitalov_krioukov};
in other words, in the previous four categories there is an implication chain that goes
from left to right, but not from right to left.

In a nutshell, 
subexponential means that the probability of the sum of two independent variables
is twice the probability of one of them;
long-tailed means that the tail is unaffected by finite shifts ($S_{long}(x+c) \rightarrow S_{long}(x)$);
and
heavy-tailed is when the tail decays more slowly than any exponential \cite{Voitalov_krioukov}.
A fundamental result of extreme-value theory is that 
fat-tailed distributions belong to the so-called Fr\'echet maximum domain of attraction, 
whereas heavy-tailed distributions that are not fat-tailed do not belong to such domain
\cite{Voitalov_krioukov}.

From an operational point of view, we define a power-law-tailed distribution
(or just a power law, pl) 
as a distribution whose tail, defined by $x$ above a lower-cut-off $u$ (i.e., $x\ge u$),
is given by the probability density
$$
f_{pl}(x) = \frac \alpha u \left(\frac u x\right)^{1+\alpha}.
$$
%above a lower cut-off $u$, that is, for $x\ge u$,
We will consider $f_{pl}(x)=0$ for $x <u$, 
that is, all empirical data below $u$ need to be disregarded \cite{Clauset}.
Note that $f_{pl}(x)$ is the minus derivative of $S_{pl}(x)=(u/x)^\alpha$ (for $x\ge u$).
Note also the difference between a fat tail, 
where the power-law may arise asymptotically, 
and a power-law tail, where the power-law arises above the cut-off $u$.
%
%A very closely related distribution is
%the Pareto (pa) distribution, 
%with density
%$$
%f_{pa}(x) = \frac \alpha \theta \left(\frac 1 {1+ x/\theta} \right)^{1+\alpha},
%$$
%for $x\ge 0$,
%with $\alpha$ the exponent (of $S_{pa}(x)$) and $\theta$ the scale parameter.
%Note that the Pareto distribution is a fat-tailed distribution, 
%and for $x\gg \theta$ it develops a power-law tail
%(in the limit $x/\theta \rightarrow \infty$, strictly speaking).
%Although there exists some confusion between the Pareto and the power-law distribution
%(in terms of their names), we take the prescription of considering the Pareto distribution
%as the distribution arising from the so-called generalized Pareto distribution
%\cite{Coles} in the case of a positive tail index $\xi=1/\alpha>0$.
%
%%coming from $S(x)=1/(1+x/\theta)^\alpha$
%
%%$$
%%f_{pa2}(x) = \frac \alpha \theta \left(\frac 1 {1+ (x-1)/\theta} \right)^{1+\alpha},
%%$$
%%for $x\ge 1$,

As an alternative description of the tail, 
we consider the example given by the truncated log-normal (ln) distribution,
which is not fat-tailed but subexponential (and therefore long-tailed and heavy-tailed
\cite{Voitalov_krioukov}).
Its probability density is
\begin{equation}
f_{ln}(x)=
{\sqrt{\frac 2\pi}}
\left[
 \mbox{erfc}\left(\frac{\ln u -\mu}{\sqrt{2} \sigma}\right)
\right]^{-1}
\frac 1{ \sigma x}
\exp\left(-\frac{(\ln x-\mu)^2}{2\sigma^2}\right),
\label{Eqlognormal}
\end{equation}
for $x\ge u$ (and zero otherwise, so the truncation is for $x<u$), 
with $\mu$ and $\sigma^2$
the mean and variance of the underlying (untruncated) normal distribution,
$u$ the lower cut-off defining the starting point of the tail,
and erfc the complementary error function \cite{Abramowitz}.

The log-normal distribution has been an important competitor of the power law
for the size distribution of structures and events in complex systems
\cite{Malevergne_Sornette_umpu,Corral_Gonzalez,Corral_Arcaute},
due to the fact that it can be described as a sort of power law
whose exponent is not constant but increases very slowly with $x$, 
when $\sigma^2$ is large.
This is commonly found in log-log plots of empirical probability densities $f(x)$,
which tend to show a slight downwards curvature
(convexity for physicists and concavity for mathematicians), 
as it would correspond to a slowly increasing exponent.
% 
%%Finally, for comparison purposes we will use 

\section{Direct fit of the tail}

We proceed to the fitting of the power-law tail and the log-normal tail to the
epidemic-size empirical data. 
Given the lower cut-off $u$, the fitting of the distributions is straightforward by maximum-likelihood estimation. 
However, the important point 
%for the power law and the log-normal 
is precisely the determination of $u$, 
both when fitting the power-law and the log-normal tail
(naturally, the value of $u$ can be different in each case, 
and we will distinguish between $u_{pl}$ and $u_{ln}$, respectively).
%(not for the Pareto distribution, whose lower cut-off is fixed to 
%%the minimum value 1000).
%zero).
Many different methods have been proposed for the determination of $u$ in the case of power-law distributions (for bibliography, see Ref. \cite{Voitalov_krioukov}),
including visual methods even \cite{Cirillo_Taleb}.

We use the fitting procedure developed in Ref. \cite{Corral_Deluca},
which is similar in spirit to the very-popular one of Ref. \cite{Clauset}
but performing much better under controlled situations \cite{corral_nuclear,Hanel_power_laws,Voitalov_krioukov}.
In essence, the procedure tries a wide range of possible values of $u$
(50 per decade, equally spaced in log-scale)
and selects the smallest one which gives an ``acceptable'' (greater than 0.20) $p-$value
in a Kolmogorov-Smirnov goodness-of-fit test.
The same method is applied to fit the truncated log-normal distribution \cite{Corral_Gonzalez}.

\begin{figure}[!ht]
\begin{center}
(a)
\includegraphics[width=10cm]{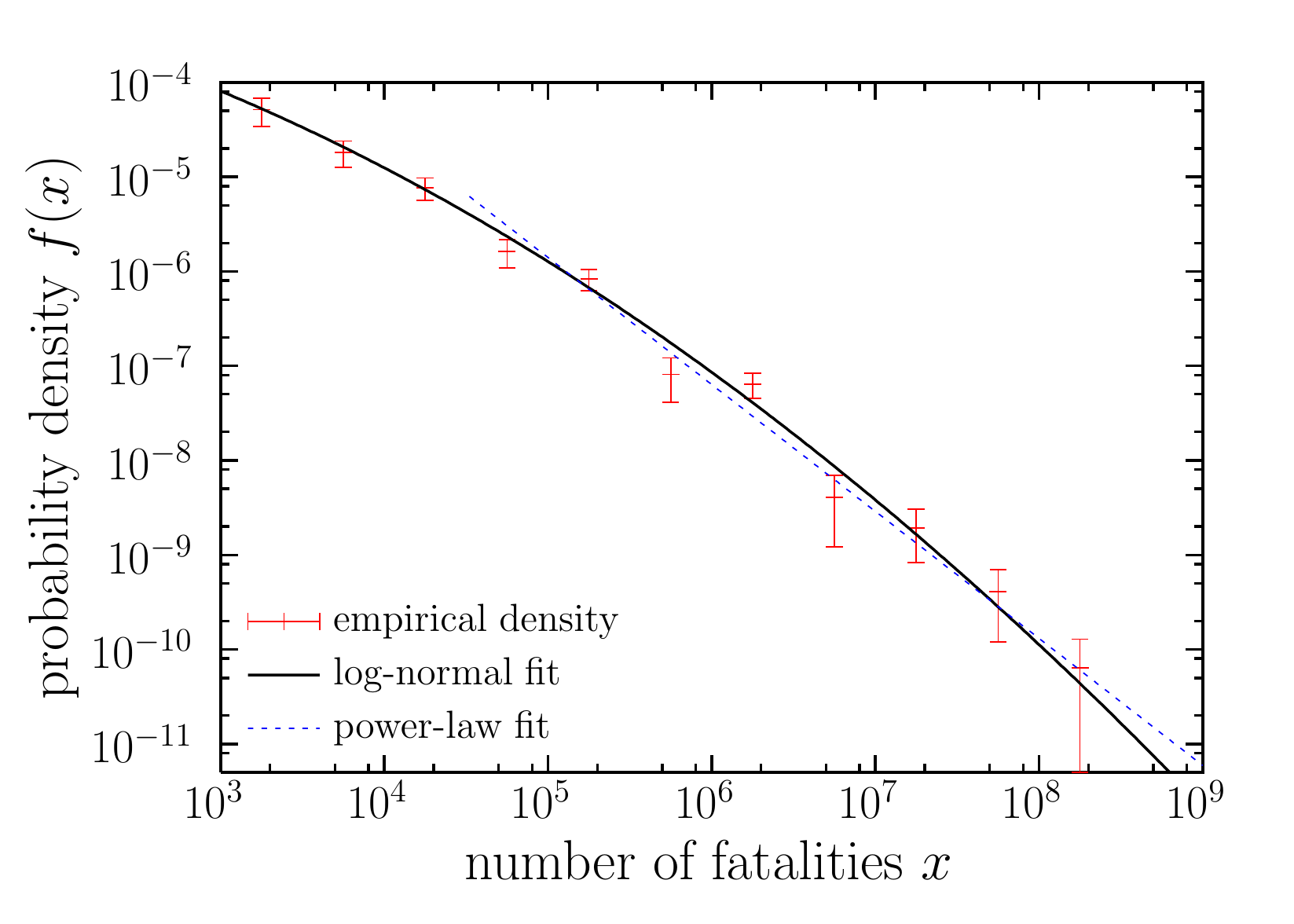}\\ %{./Fig_density.pdf}
(b)
\includegraphics[width=10cm]{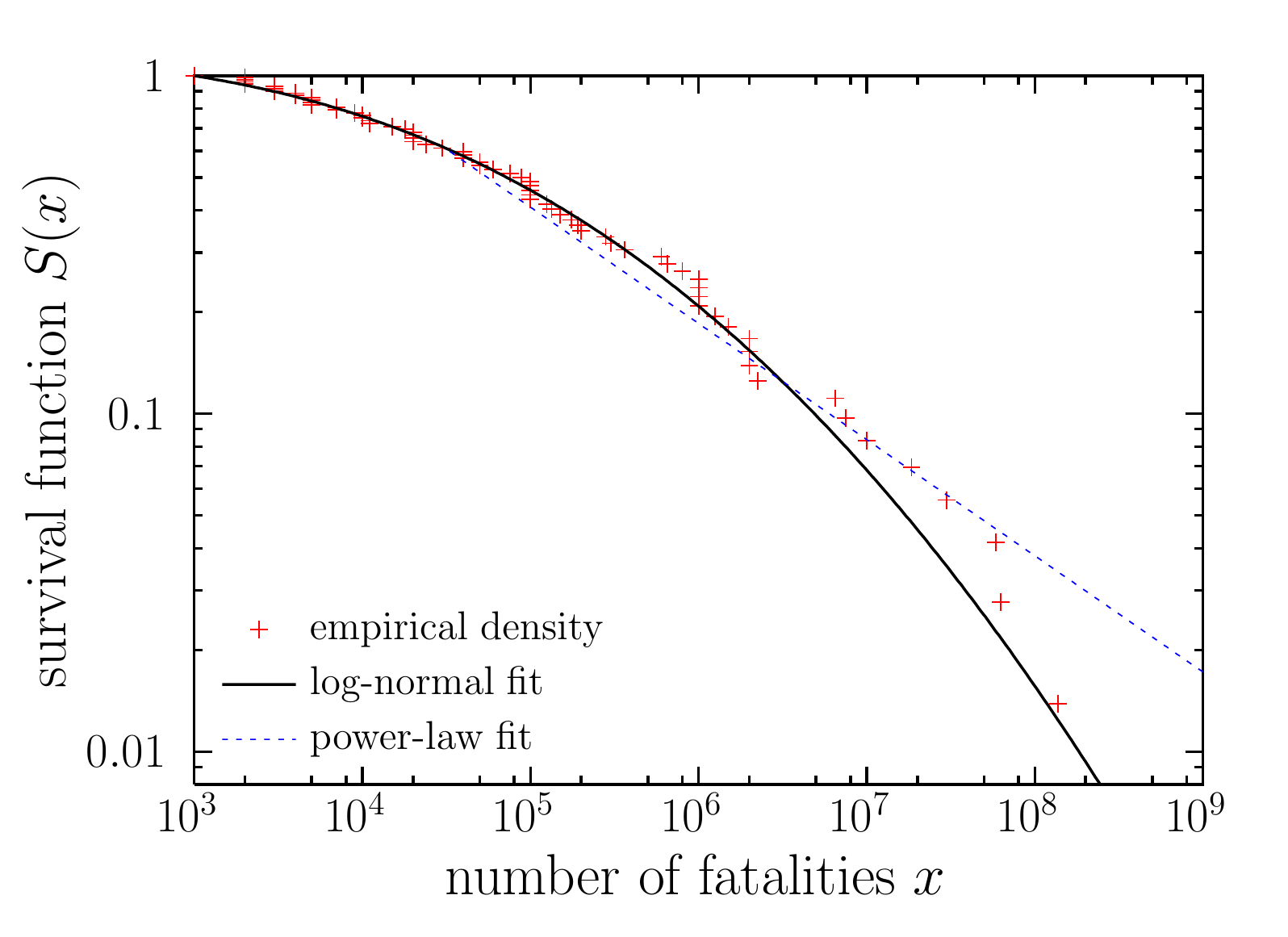} %{./Fig_density.pdf}
\end{center}
\caption{
Empirical distribution
of the number of fatalities for each of the 72 historical epidemics
in the original data set
studied in Ref. \cite{Cirillo_Taleb}.
A truncated log-normal fit and a power-law tail (starting at $u_{pl}\simeq 33,000$)
are shown as well.
(a) Probability density
 (empirical distribution
obtained using logarithmic binning \cite{Corral_Deluca}).
(b) Complementary cumulative distribution function
(i.e., survival function).}
\label{Fig2}
\end{figure}

%We fit a truncated log-normal tail to the epidemic data of Ref. \cite{Cirillo_Taleb}
%using the method of Ref. \cite{Corral_Gonzalez} to find the lower cut-off $u$, 
A visualization of the resulting fits in comparison with the empirical estimations of the
probability density 
%of the number of fatalities 
(using logarithmic binning \cite{Corral_Deluca}) 
and of the survival function
is provided in Fig. \ref{Fig2}, using the original data set.
It is apparent how
both fits are very close to each other
(the difference being much smaller than the uncertainty in the empirical values of the density)
and also that the power law
gives a probability higher than the log-normal for the most extreme events (as expected).
The log-normal fit yields
$u_{ln}= 1000$, %(fitting the whole data set),
$\mu=10.43$ %3.56+dlog(1000.d0) ! fits a valores originales (sin multip X 1000)
and
$\sigma=3.60$
%$\mu=10.4$ % valor corregido ??
%and $\sigma=3.60$
($p-$value $0.94$; scale parameter $e^\mu\simeq 35,200$),
%The empirical estimation of the probability density of the data
%%using logarithmic binning \cite{Corral_Deluca}
%together with the obtained fit
%are shown in Fig. \ref{Fig2}.
%leading to the parameters
%It is remarkable that the power-law fit yields $u\simeq 33,000$
%and $\alpha=0.344\pm0.005$ (one standard deviation),
%whereas the log-normal yields 
%$u= 1000$ (fitting the whole data set)
%and $\mu=10.4$ % valor corregido, reescalado factor 1000!! Original 3.56
%and $\sigma=3.60$
%(scale parameter $e^\mu\simeq 34,000$).
%see Table \ref{table1}.
%The figure %sets clearly that, over the common range, 
%which also compares 
%A comparison 
%with a power-law (pl) fit for the tail 
%%($f_{pl}(x) \propto 1/x^{1+\alpha}$, see Ref. \cite{Corral_Gonzalez})
%shows that
%both fits are very close to each other
%(the difference is much smaller than the uncertainty in the empirical value),
whereas for the power law
%but also that the power law, with 
$u_{pl}\simeq 33,000$ and $\alpha=0.34$
($p-$value $0.21$), see Table \ref{table1}. 
%overestimates the probability of large events, in comparison with the log-normal.
%gives a higher probability than the log-normal for the most extreme events (as expected).

%FALTA PARETO FIT HERE!!
%
The results %of the power-law fit  
for the restricted data set (years 1485-2000) are comparable.
The graphical estimations of the distributions (not shown)
look rather like the previous case.
The power-law fit yields
a close value of $\alpha$, %0.325
but a smaller $u$, as expected ($u_{pl}\simeq 14,000$), 
due to less undersampling of small $x$ 
(Table \ref{table1}). % \ref{table1}.
The log-normal fit also leads to values of the parameters close to those
of the original data set, as also shown in the table. % \ref{table1}.
In addition, for the rescaled data set we find the same overall behavior
(see the table once more).
Thus, we conclude that the properties of
the epidemic-size distribution
are roughly the same for the original data set,
for the restricted data set, 
and for the rescaled data set.
This does not mean that incompleteness is not relevant, 
only that the degree of incompleteness is similar for
the three data sets. 
%Further, 
%the main properties 
%of the tail of the epidemic-size distribution
%do not depend much neither on its incompleteness
%nor on the time inhomogeneity of total world population.

In this way, for the rest of the paper we concentrate on the original (not rescaled) data set 
for a more in-depth analysis,
which will
compare the statistical behavior of this empirical data with that of 
many realizations of the simulation
of a truncated log-normal distribution, 
with $N=72$ events and the values of the parameters 
$u_{ln}$, $\mu$, and $\sigma$ given above 
(and also in the first row of Table \ref{table1}).

%MENTION here that we do may realizations!!!!

One could argue that it would be fair to compare the fit of the truncated log-normal (which has finite mean, and is truncated from below) 
not with a power-law tail (with infinite mean)
but with the ad-hoc tapering of the power-law tail used in Ref. \cite{Cirillo_Taleb}.
We argue that there is no practical difference between both types of power laws
in relation to the fitting of the empirical data.

Indeed, the authors of Ref. \cite{Cirillo_Taleb} introduced the change of variable
%$$
%x'=h-(h-l) \exp\left(-\frac{x-l}\Delta\right),
%$$
%$$
%x'=h-(h-l) e^{-(x-l)/\Delta},
%$$
\begin{equation}
z=l-\Delta\ln \left(1-\frac{x-l}{h-l}\right)
\end{equation}
with $h$ the total world population, 
$l$ the smallest value taken by $x$,
and
$\Delta=h-l\simeq h$.
The key point is that 
when $x=h$ then $z\rightarrow \infty$,
whereas when $x\ll h$ then $x\simeq z$,
and thus, a power law (untruncated from above) for $z$ corresponds to
a tapered power law for $x$, with a ``sharp'' truncation at $x=h$
(for what a strange thing a ``sharp'' truncation can be, see Ref. \cite{Corral_comment_CT2}).

The authors of Ref. \cite{Cirillo_Taleb} propose to transform the original data $x$
into $z$, fit an power-law tail $f_{pl}(z)$ to $z$,
and then transform back to obtain the tapered power-law (tpl) fit $f_{tpl}(x)$ of $x$.  
The resulting tpl distribution for $x$ is given by
$f_{tpl}(x)=f_{pl}(z)dz/dx$, with $dz/dx=\Delta/(h-x)$.
Taking values $h=7.7\times 10^9$ 
(current world population, as suggested in Ref. \cite{Cirillo_Taleb})
and $l=1000$, we get that the largest change in $x$ is for the largest $x$,
which is $x=1.375 \times 10^8$,
resulting in
$z=1.387 \times 10^8$ 
(and $dz/dx=1.018$).
Thus, the changes are so small that 
it yields the same results to fit a power law to the empirical data $x$ or to 
the transformed data $z$,
and then the fit given by $f_{tpl}(x)$
is, in practice, indistinguishable from what one obtains fitting directly the untruncated power law $f_{pl}(x)$ to $x$.

%ESTO HAY QUE EXPLICARLO MEJOR!!

%MENCIONAR TAMBIEN EL FIT LOGNORMAL ARRIBA??

%
%Next we will compare the statistical behavior of the epidemic empirical data
%of Ref. \cite{Cirillo_Taleb} with that of the simulation
%of a truncated log-normal distribution (with the values of the parameters given above
%and $N=72$ events).
%

%DECIR QUE EL NOMBRE SURVIVAL ES CONFUSO??

{
\begin{table}[h]
%\begin{centering}
\begin{center}
\caption{\label{table1} 
Results from fitting a power-law tail (pl) and a truncated log-normal (ln)
to the three data sets under consideration: 
(1) original, 
(2) restricted to the period 1485-2000 AD, 
and 
(3) original but rescaled by world population.
$x$ is number of fatalities except for (3), where it yields fatalities divided by world population at the time of the epidemic;
$x_{max}$ is the largest value of $x$ on record;
OM is the number of orders of magnitude covered by each fit, 
calculated as $\log_{10} (x_{max}/u)$;
$n$ is the number of events in the fitting range (above $u$);
and
$p$ is the $p-$value of the fit.
The results for the largest values of $u$ 
for which the log-normal fit does not bring any improvement 
with respect the power-law fit, given by the log-CV test (cv),
are also included. 
}
\smallskip
% latex_power_law_notru_col4_generic_epidemic.txt                              
\begin{tabular}{ |l|cc | cc ccc | ccc ccc|ccc|}
%data    
\hline
& $N$ & $x_{max}$ & $u_{pl}$ & OM$_{pl}$ & $n_{pl}$ & $\alpha_{pl}$ & $p_{pl}$ & $u_{ln}$ & OM$_{ln}$ & $n_{ln}$ & $\mu$ & $\sigma$ & $p_{ln}$ &$u_{cv}$ & OM$_{cv}$ &$n_{cv}$\\
\hline    
1 %full 
&        72 &    $138\times 10^6$ &   33,100 &      3.6 &        43 &        0.344           &  0.20 & %%\\%                           cirillo_taleb_data_1000
   1000 &    5.1 &        72 &       \phantom{$-$}10.43     &   3.60  &  0.94 & %%\\%                           cirillo_taleb_data_1000
364,000 & 2.6 & 21\\ 
%%%%%%%%%%%%%%%%
2 %1485-2000 
&        50 &   $\phantom{1}59 \times 10^6$ &   13,800 &      3.6 &        38 &        0.325           &  0.22 & %%\\%                      cirillo_taleb_data_1485_2000 % corregido X 1000
    1000 &    4.8 &        50 &        \phantom{$-$}11.01     &   3.13  &  0.85 &%%\\%                      cirillo_taleb_data_1485_2000
600,000 &2.0 &15\\
%%%%%%%%
3 %full, rescaled 
&        72 &   0.35 &   $5.2\times 10^{-5}$ &      3.8 &        42 &        0.351          &  0.21 & %%\\%                            cirillo_taleb_data_sca
    $2.5\times 10^{-7}$ &    6.1 &        72 &       $-10.10$       & 4.04  &  0.41 &%%\\%                           cirillo_taleb_data_sca2
$2.7\times 10^{-4}$ & 3.1&28\\
%%pl &        72 &   .832E+07 &   .138E+01 &      6.780 &        61 &        0.311         .041  &  .20\\%                                          _simu_PL
%%Par &        72 &   .250E+09 &   .120E+02 &      7.317 &        61 &        0.237         .031  &  .20\\%                                      _simu_Pareto
%\hline
%ln1 &        72 &   .180E+09 &   .120E+05 &      4.2 &        53 &        0.299        &  .25 &%%\\%                                     sim_lognormal
\hline
\end{tabular}
\par
\end{center}
%\centering{}%\caption{}
\end{table}
}

\section{Mean-excess size and maximum-to-sum ratio}

\subsection{Mean-excess size}

Reference \cite{Cirillo_Taleb} proposes two main ways to check fat-tailness.
One of them uses
the mean-excess function $\epsilon(u)=\langle x-u | x\ge u \rangle$, 
where the brackets denote expected value
and the vertical bar denotes conditioning.
This is the same as the expected residual size
\cite{Kalbfleisch2} (see also Ref. \cite{Schroeder}) 
used in reliability theory and survival analysis,
which characterizes a probability distribution in a way totally equivalent
to $f(x)$ or $S(x)$, provided that the first moment of the distribution 
($\langle x \rangle=\epsilon(0)$) is finite, which for a power-law tail 
happens when $\alpha > 1$.
In this case, $\epsilon(u)$ should increase linearly with the lower cut-off $u$.

However, for a power-law tail with $\alpha < 1$ 
the first moment does not exist (is infinite),
which implies that the %(theoretical) 
mean-excess function does not exist either.
Thus, in this case, 
the attempt of direct empirical estimation of $\epsilon(u)$ used in Ref. \cite{Cirillo_Taleb}
(replacing the expected value in $\epsilon(u)$ by the sum for the empirical values)
is futile, 
as one cannot estimate something that does not exist.
Therefore, the empirical results of Ref. \cite{Cirillo_Taleb}
%and then its estimation from data cannot converge.
%This makes the use of this method in Ref. \cite{Cirillo_Taleb} 
are void of theoretical support in the case of power-law tails
(which is the case considered in that reference).
%In other words, one can emulate an estimator of $\epsilon(u)$, 
%replacing the expected value by the sum for the empirical values

\begin{figure}[!ht]
\begin{center}
(a)
\includegraphics[width=7.5cm]{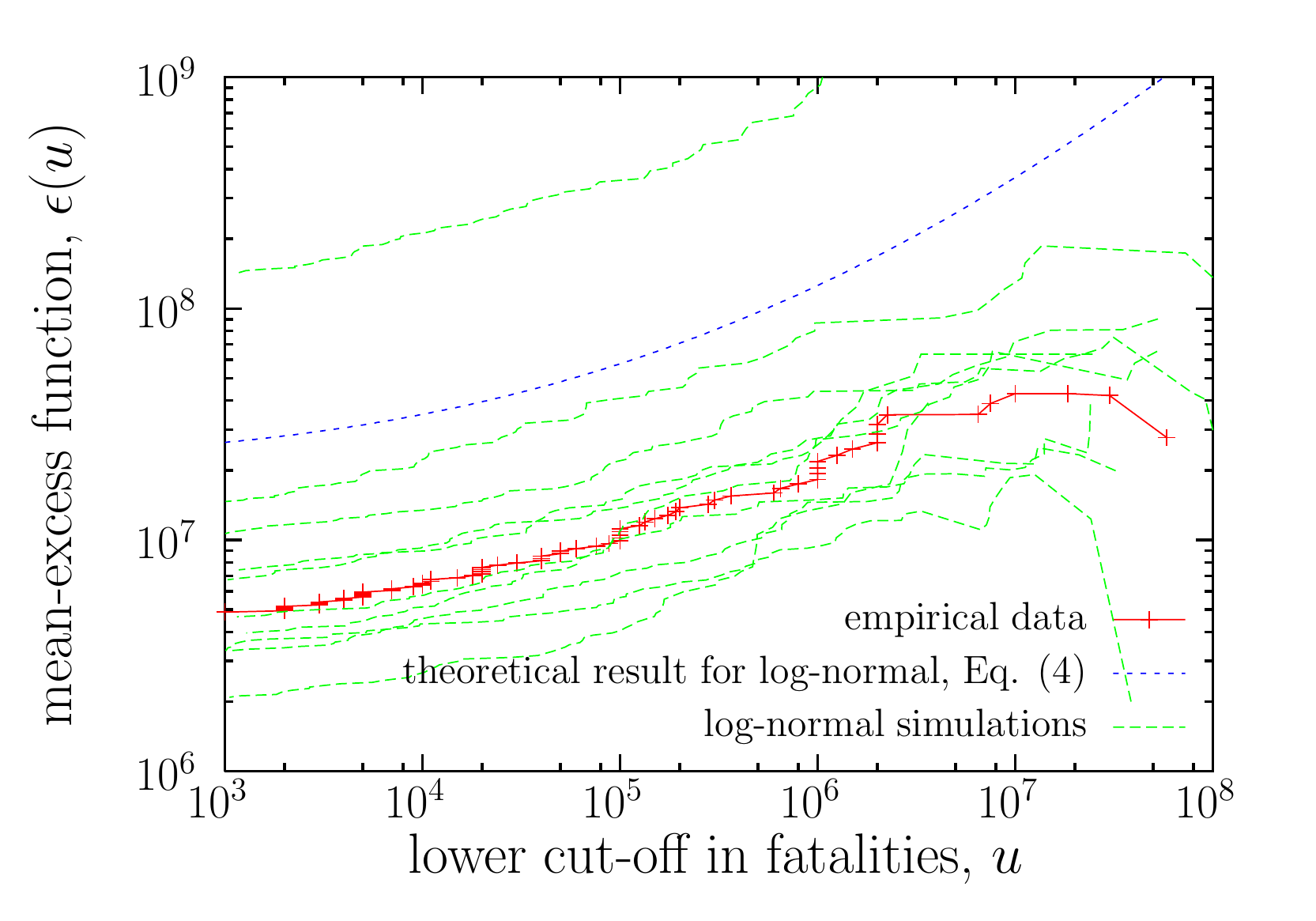}%{./Fig_excess.pdf}
(b)
\includegraphics[width=7.5cm]{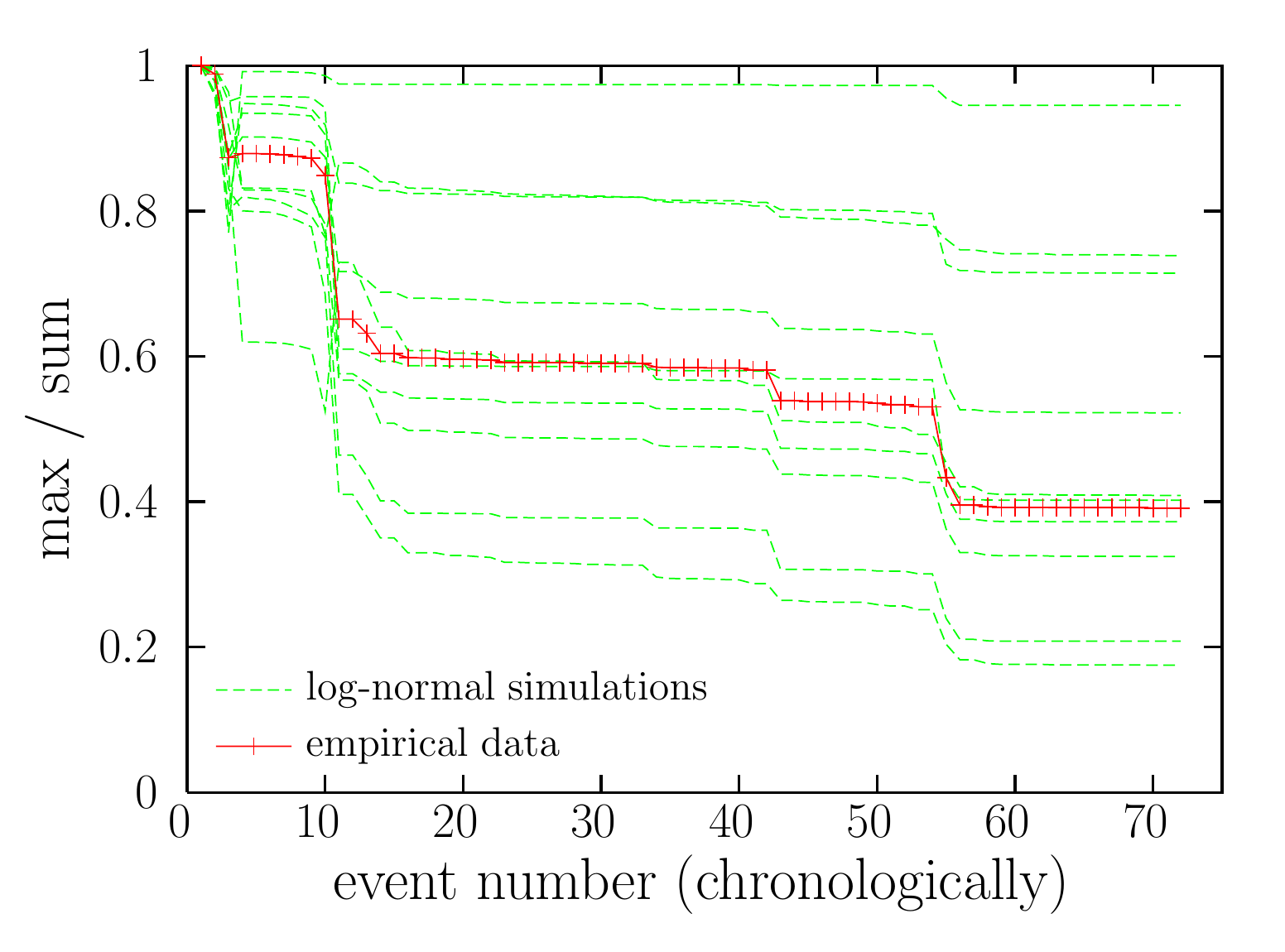}\\%{./Fig_max_to_sum.pdf}\\
(c)
\includegraphics[width=7.5cm]{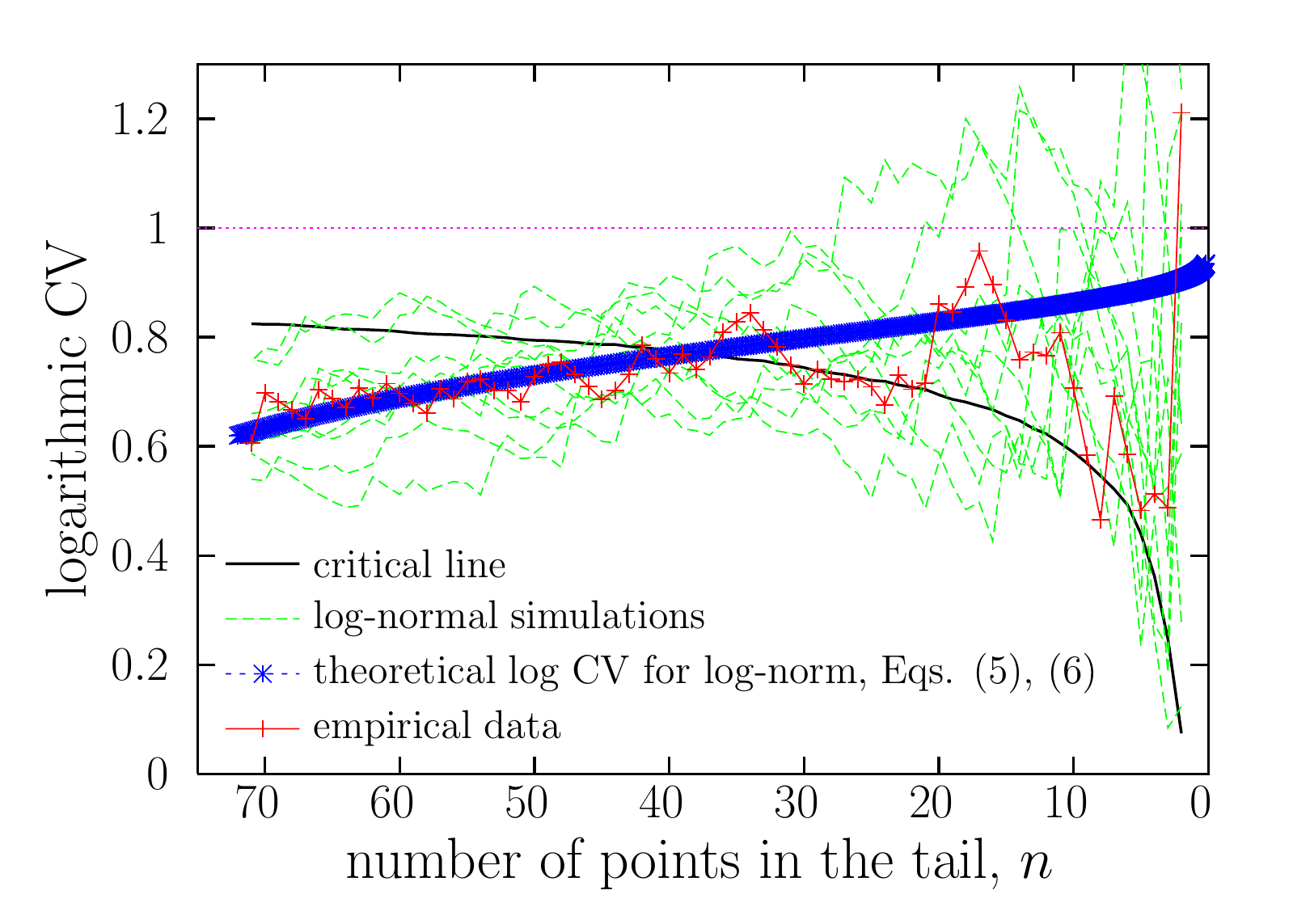}
(d)
\includegraphics[width=7.5cm]{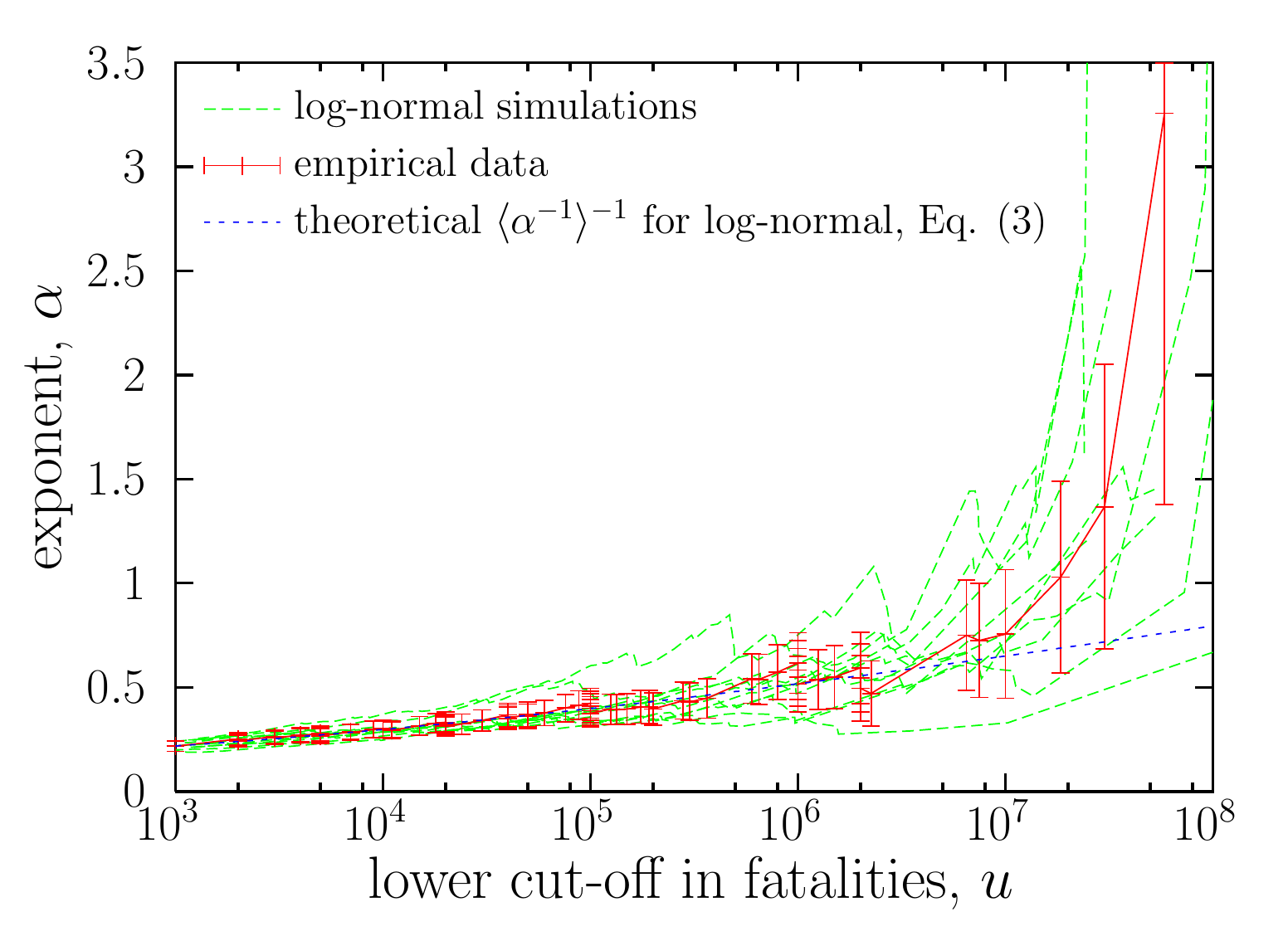}\\%{./Fig4expo_cutoff_a.pdf}
\end{center}
\caption{
Statistical properties of the number of fatalities of historic epidemics
compared to those of 10 truncated log-normal synthetic samples
with the parameters that fit the empirical data 
(Table \ref{table1} and Fig. \ref{Fig2}).
(a) Estimation of mean-excess function versus minimum size (lower cut-off) $u$;
the theoretical calculation for the fitted log-normal is also shown.
(b) Maximum-to-sum ratio as a function of number of data, 
in chronological order.
(c) Logarithmic coefficient of variation as a function of the number of points in the tail
(those with $x > u$; note that the horizontal axis is reversed).
%DAR LOS VALORES CRITICOS!!!!!
The critical line corresponds to the 5th percentile, 
so the significance of the test is 0.05.
The theoretical value of the logarithmic CV for the truncated log-normal is also shown.
(d) Estimated power-law exponent $\alpha$ as a function of $u$,
the error bars ($\alpha/\sqrt{n}$) denote one standard deviation;
the inverse of the expected value of $\alpha^{-1}$
calculated for truncated log-normally distributed data is also shown.
}
\label{Fig5}
\end{figure}

Figure \ref{Fig5}(a) shows
%of the three simulated distributions (power law, log-normal, and Pareto),
%the mean-excess function of the empirical data, 
this empirical estimation for the original data set of epidemic fatalities \cite{Cirillo_Taleb}
and compares with the estimation of the mean-excess function 
%and how this behaves in the same way as
%one realization of % esto explicarlo mejor en un apendice?? 
for several realizations of log-normally simulated data 
(with the parameters obtained from the log-normal fit, Table \ref{table1}).
Although there is considerable dispersion in the different realizations, 
there is no way to distinguish the empirical data
from the log-normal simulations.
Moreover, note that
some degree of concavity (convexity for mathematicians) of the 
log-log plot seems to indicate that the empirical function would
increase with $u$ faster than any power of $u$, 
and thus faster than linearly (except for the four most extreme events).
%is rather close to the 
%shows the same pattern as the

\subsection{Maximum-to-sum ratio}

The other approach in Ref. \cite{Cirillo_Taleb} to fat-tailness
uses the (partial) maximum-to-sum ratio
(the maximum of the $x-$values divided by the sum of the values),
with the data swept in chronological order.
For $N\rightarrow \infty$ this ratio should tend to zero 
when the mean of the distribution is finite 
(as it happens with the log-normal 
but not with the power law when $\alpha < 1$, 
because both the maximum and the sum of a power law with $\alpha<1$ 
scale superlinearly, 
as $N^{1/\alpha}$ \cite{Bouchaud_Georges,Corral_csf}). 

We again compare the empirical data with the log-normal simulated data, 
sorting the simulated data in order that the ranks of the sizes 
(number of fatalities)
follow the same temporal pattern as the empirical data 
(i.e., the largest simulated event is %always 
put on the 11th position, where the Black Death,
the largest event on record, takes place in the original data, and so on).

The results, displayed in Fig. \ref{Fig5}(b), 
show again that 
the behavior of 
the epidemic empirical data
is indistinguishable from 
what is obtained for 
synthetic log-normal data (with a remarkable scattering),
for which the max-to-sum ratio decreases with the number of data 
but without reaching zero.
Thus, although the theory teaches us that the maximum-to-sum ratio tends to zero 
when $N\rightarrow \infty$ if the distribution has a finite mean, 
this convergence can be rather slow, 
as it happens with the log-normal distribution for the parameter values
that describe the epidemic data.
%for which the maximum-to-sum ratio decreases but does not tend to zero,
%showing the same behavior as the empirical data.

In summary, the two methods used in Ref. \cite{Cirillo_Taleb}
to provide evidence of fat-tailness for the description of the epidemic data
are not able to rule out a log-normal tail. 
%for the description of the epidemic data.

%???
 
%\begin{figure}[!ht]
%\begin{center}
%\includegraphics[width=\textwidth]{./Fig_max_to_sum.pdf}
%\end{center}
%\caption{
%}
%\label{Fig6}
%\end{figure}

\section{Logarithmic-CV test and change of apparent exponent with cut-off}

\subsection{Logarithmic coefficient-of-variation test}

Now we expose complementary evidence that the truncated log-normal distribution 
can provide a good description for the tail of the epidemic data of Ref. \cite{Cirillo_Taleb}.
In fact,
%Which of the distributions best fits the data?
%On the one hand, 
the power law can be considered a
particular case of the truncated log-normal
(in the same way that the exponential is a particular case of the truncated normal distribution when $\sigma^2 \rightarrow \infty$ and $\mu \rightarrow -\infty$
with $\mu/\sigma^2$ constant
\cite{Castillo}). 
%% $\mu - \ln u \rightarrow -\infty$ ??
In this sense, a log-normal tail cannot provide a worse fit
than a power law (it will be the same, or better).
However, on the other hand, 
it may happen that this improvement in the fit is not significant,
and then the power-law tail suffices for describing the data (as it has one parameter less that the log-normal). 
This is something that can be evaluated by a likelihood-ratio (LR) test \cite{pawitan2001}.

Taking advantage of the fact that the LR between both distributions
is a decreasing function of the logarithmic coefficient of variation 
(logarithmic CV) \cite{Castillo},
this provides a very simple way to perform the LR test
(without the need of performing maximum-likelihood fitting):
critical values of the LR translate into critical values of the logarithmic CV.
When this quantity is close enough to one, 
the test fails to reject the more parsimonious power-law hypothesis
(although the log-normal is not rejected, but the gain it brings in likelihood 
is ``superfluous'').
%the power-law hypothesis cannot be rejected, 
When the LR departs significantly from one (from below),
the power law is rejected in favor of the log-normal.  
This simple procedure provides a
uniformly most powerful unbiased test 
between 
a power-law tail (null hypothesis $\mathcal{H}_0$)
and a log-normal tail (alternative hypothesis $\mathcal{H}_1$, 
containing $\mathcal{H}_0$ as a special case)
\cite{Castillo,Malevergne_Sornette_umpu}.
%for a log-normal tail in front of a power-law tail.

The test is performed for different values of the lower cut-off $u$, 
and the results, for the original data set, are displayed in Fig. \ref{Fig5}(c).
This shows that for the 21 largest epidemics, the power law tail is preferred
(although remember that the log-normal tail is not rejected, 
as, being a generalization of the power law, with an additional parameter, 
cannot yield a worse fit).
When the tail comprises more than 21 epidemics,
the power-law is rejected in favor of the log-normal tail.
The corresponding value of the cut-off $u$ for this transition, $u_{cv}$, 
turns out to be at about 360,000 fatalities (Table \ref{table1}).
In other words, the 21 epidemics with more than 360,000 fatalities are well described
by a power law (the improvement brought by the log-normal is not significant), 
but, including events below 360,000, the log-normal fit is significantly better
(for the full range).

Applying the same procedure to the log-normally simulated data
yields the same conclusion as in the previous section:
the pattern obtained for the empirical data is indistinguishable from log-normal simulations,
as also shown in Fig. \ref{Fig5}(c).
It is relevant to notice how, for log-normally distributed data 
(at least for the parameters in Table \ref{table1}, first row),
there exists a tail 
that is indistinguishable from a power-law, 
and therefore the power-law is preferred, despite the simulated data are log-normal
(by construction).
This is an unavoidable consequence of the similarity between truncated log-normals 
and power laws, and implies that the fact that for the 21 largest epidemics the preferred fit is power law does not guarantee that the underlying distribution is not log-normal.
%The results for the three data sets are also included on Table \ref{table1}.

Note that
in the case of the log-normally simulated data, as in the previous section, 
there is substantial scattering in the simulations. 
For the particular simulations shown in Fig \ref{Fig5}(c)
the range of points in the tail for which the power law is not rejected
ranges from 3 to 68, with an average equal to 23
(with all cases of log-normally simulated data 
displaying a non-rejectable power-law tail). 
In addition, the logarithmic coefficient of variation of a truncated log-normal
random variable can be exactly calculated 
(see the Appendix I, Eqs. (\ref{finalC}) and (\ref{finalC2})).
This theoretical value is also shown in Fig. \ref{Fig5}(c), using the fitted values of $\mu$ and $\sigma$ to parameterize it.
The agreement with the empirical data is remarkable,
showing that the log-normal gives a better representation of the epidemic data
than a simple power-law tail
(and, as expected, the dispersion of the log-normal simulations is centered around the
theoretical value). 

\subsection{Drift of the apparent power-law exponent}

Assuming that a power law could describe the largest epidemics (in terms of fatalities),
which would be the value of the corresponding power-law exponent $\alpha$?
Above, for $u\simeq 33,000$, we report $\alpha=0.34$,
but for $u\simeq 360,000$ the maximum-likelihood estimation turns out to be larger: 
$\alpha=0.45$.
% columna 1 y 13 de
% C:\Users\acorr\OneDrive - Universitat Autònoma de Barcelona\projects\epidemic\cvtest\cvtest
In fact, the estimated value of $\alpha$ is not stable at all, 
growing when the lower cut-off $u$ increases,
reaching $\alpha>1$ for the highest values of $u$, see Fig. \ref{Fig5}(d);
that is, the fatness of the tail $\xi=1/\alpha$ decreases systematically with $u$
%(see Fig. \ref{Fig5}(d),
(this is
already apparent in one of the plots of Ref. \cite{Cirillo_Taleb}), 
which prevents that one can establish a well-defined exponent
\cite{Baro_Vives}.
This increase of the apparent exponent $\alpha$ beyond one
seems to contradict 
fat-tailness in general and
the ``extreme fat-tailness'' claimed in Ref. \cite{Cirillo_Taleb} in particular.

Additionally,
extreme-value theory \cite{Coles}
ensures that, for asymptotically large thresholds $u$
(when the values of $x$ are independent and identically distributed), 
the probability distribution of threshold exceedances $x-u$
tends to a generalized Pareto distribution (GPD).
The parameter $\xi$ of the GPD separates fat (power-law) tails
(for $\xi>0$, corresponding to the Fr\'echet maximum domain of attraction) from other tails ($\xi=0$ or $\xi<0$).
But notice that, although $\xi$ can be estimated from data using the Hill estimator
(which is totally equivalent to the maximum-likelihood estimator for $\alpha$ we have used),
%Notice that even though $\xi$ is the parameter separating  
%fat tails from light (exponential) tails in the generalized Pareto distribution,  
the fact of obtaining a positive value of $\xi$
does not mean that the data are fat tailed. 
The reason is that this calculation of $\xi$ (or $\alpha$)
%by maximum-likelihood estimation 
assumes that $\xi$ is positive.
In other words, one can never obtain a zero or negative value of $\xi$ from the Hill estimator, which assumes fat-tailness by construction.
So, contrary to what one may think, 
the positive values of $\xi$ obtained in Ref. \cite{Cirillo_Taleb},
and also here (Fig. \ref{Fig5}(d)), do not provide evidence of fat-tailness. 
%This inconvenience, however, can be useful to give further evidence about
%which is the best distribution to describe the data.
%
%We simulate $N=72$ values of a random variable for 
%the three different theoretical distributions under consideration:
%power law, Pareto, and log-normal.
%We use as input parameters the ones obtained from the fits of the complete empirical data 
%(Table \ref{table1}, first row),
%and fit power-law distributions to the simulated data
%for different values of the lower cut-off $u$.
%Indeed, 

Figure \ref{Fig5}(d), 
in addition of showing the resulting exponents $\alpha$ as a function of $u$ for the empirical data,
also compares with the log-normally simulated ones.
Once more, 
it is clear that the simulated data provides a pattern compatible with 
the empirical one, with an increase of the value of the exponent $\alpha$
when $u$ increases (and with positive $\xi$, despite the log-normal does not belong
to the Fr\'echet maximum domain of attraction). 
Indeed, this increasing behavior of the fitted exponent is what one expects from 
a log-normal distribution, for which
the probability density appears as ``convex'' in a log-log plot.
A power-law with $\alpha<1$ does not provide such a systematic increase of 
$\alpha$, in general.
%In contrast, the power-law simulated data is not able to replicate the empirical pattern so well as the log-normal and the Pareto.
%Performing the logarithmic CV test for the simulated data 
%we arrive to similar conclusions, see Fig. \ref{Fig3}.

%OJO AL DATO, CRECE LINEALMENTE EN LIN-LOG PLOT, 
%INDICATIVO DE UN LOGARITMO...
%CUADRA CON LOGNORMAL???

We can go one step forward and provide a theoretical calculation of the value of $\alpha$ resulting from the fitting of a power-law tail to truncated log-normal data,
in a spirit similar to that of Ref. \cite{Salje_Vives}.
The maximum likelihood estimation of $\alpha$ is simply the inverse of the sample mean of $\ln x - \ln u$ \cite{Corral_Deluca}. 
Assuming than $x$ follows a truncated log-normal for $x\ge u$, with parameters 
$\mu$ and $\sigma^2$, 
then $\ln x - \ln u$ follows a truncated normal distribution for $\ln x - \ln u \ge 0$,
with parameters 
$\tilde\mu=
\mu-\ln u$ and $\sigma^2$.
The expected value of such distribution gives, therefore, the expected value of the inverse of the estimation of the exponent (or the expected value of $\xi$), 
and can be easily calculated to be
\begin{equation}
\left \langle \frac 1 \alpha \right\rangle_{ln} = 
\langle \xi \rangle_{ln} =
\mu-\ln u+
\sigma \sqrt{\frac 2 \pi}
\,
\frac{e^{-(\ln u-\mu)^2/(2\sigma^2)}}{\mbox{erfc}[(\ln u-\mu)/(\sqrt 2 \, \sigma)]}
%\frac{e^{-\mu'^2/(2\sigma^2)}}{\mbox{erfc}[-\mu'/(\sqrt 2 \, \sigma)]}
%%\frac{e^{-\mu'^2/(2\sigma^2)}}{1-\mbox{erf}[-\mu'/(\sqrt 2 \, \sigma)]}.
\label{xilognormal}
\end{equation}
% https://en.wikipedia.org/wiki/Truncated_normal_distribution
[obtained also by direct integration of $(\ln x - \ln u)f_{ln}(x)$, from Eq. (\ref{Eqlognormal}), see the Appendix I].
Figure \ref{Fig5}(d) includes a comparison 
between 
the formula for $\langle \alpha^{-1}\rangle_{ln}^{-1}$
and
the empirical estimation of $\alpha$, as a function of $u$.
The nearly perfect agreement between both is an extra argument in support
of the plausibility of the truncated log-normality of the epidemic data.

\section{Expected final size of the current COVID-19 pandemic}

%CORREGIR ESTIMACION!!!!

Having in mind the limitations of the current study
(incompleteness of historical records, 
information limited to just one value of a random variable per epidemic,
mixture of totally different historical periods for the characteristics of epidemics...),
an estimation of the expected final death toll from the current COVID-19 pandemic
has to be understood only as a mathematical exercise.
Nevertheless, 
this exercise can be very illuminating to learn about the counterintuitive
properties of fat-tailed processes. % (fat-tailed or not).

The question is:
given that (at the time of the first submission of this article) 
the number of fatalities of the COVID-19 pandemic is $u\simeq 1,000,000$,
which is the value we can expect for its (final) size?
%which can we expect its final size will be?
%is the expected value of its final size?
What we are asking for is $\langle x | x \ge u\rangle$, 
which is directly related to the 
mean-excess function %(or expected residual size) 
$\epsilon(u)$ by
$\langle x | x \ge u\rangle = \epsilon(u)+u$.
The empirical estimation shown in Fig. \ref{Fig5}(a)
allows a direct calculation of this, 
turning out to be
$\langle x | x \ge 10^6\rangle \simeq
2.1 \times 10^7$.
This is based just on the 18 historic events with $x \ge 10^6$. 

We can try to use the diverse theoretical distributions, 
as arising from the fits, to try to improve this estimation.
For a power-law distribution, we know that 
$\langle x | x \ge u\rangle_{pl} = \alpha u / (\alpha-1)$;
%If we assume a value of $\alpha=1.1$, this leads to 
%$\langle x | x \ge u\rangle_{pl} = 11u \simeq ???$,
%i.e., larger than a 1000 \% increase of the current figure $u$.
%If we assume $\alpha=0.9$, the resulting expected value becomes infinite,
%which is true for any $\alpha < 1$.
however, as we have already mentioned,
this is only valid for $\alpha >1$ \cite{Schroeder}.
For values of $\alpha$ below one (proposed in Ref. \cite{Cirillo_Taleb}),
$\langle x | x \ge u\rangle_{pl}$ becomes infinite
(although finite for an estimation of a finite sample
and strongly dependent on $N$).
%The Pareto distribution also leads to the same result,
%$\langle x | x \ge u\rangle_{pa} \infty$.
In contrast, for the log-normal distribution, 
we can calculate the expected value from simulations
(the existence of $\langle x | x \ge u\rangle_{ln}$ ensures covergence,
in contrast to the power-law case with $\alpha<1$);
nevertheless, the convergence is rather slow, for $N=10^6$ we obtain
$\langle x | x \ge 10^6\rangle_{ln} \simeq
(1.09 \pm 0.04) \times 10^8$,   % con N=1e6 
%25,000,000$,  % de una realizacion con N=72 ! too low!!!!
% la media de la distribucion (con u=1000) sale 2.2e7 con N=1e6 (Taleb 7 millones)
% y para u=200000 sale 6.1e7 con N=1e6 (mucho mas grande que Taleb, que da 20.1 millones)
which seems exaggeratedly large, but at least it is not infinite.
In any case, this calculation clearly demonstrates that
the risk posed by a log-normal tail should not be disregarded. 

In fact, $\langle x | x \ge u\rangle$ can be calculated analytically for the 
truncated log-normal distribution, starting from 
$\langle x | x \ge u\rangle = \int_u^\infty dx\, x f(x)/S(u)$.
Let us denote $f_{ln}(x)=f(x;\mu,\sigma^2)$, then
$$
x f(x;\mu,\sigma^2) = f(x;\mu+\sigma^2,\sigma^2) \,
\frac{\mbox{erfc}\left(\frac{\ln u-\mu-\sigma^2}{\sqrt{2}\sigma}\right)}
       {\mbox{erfc}\left(\frac{\ln u-\mu}               {\sqrt{2}\sigma}\right)}
\,
e^{\mu+\sigma^2/2}.
$$
Identifying $u$ with the lower cut-off of the distribution, $S(u)=1$, 
and taking advantage that $f(x;\mu+\sigma^2,\sigma^2)$ has to be normalized
we obtain
\begin{equation}
\langle x | x \ge u\rangle_{ln}=
\frac{\mbox{erfc}\left(\frac{\ln u-\mu-\sigma^2}{\sqrt{2}\sigma}\right)}
       {\mbox{erfc}\left(\frac{\ln u-\mu}               {\sqrt{2}\sigma}\right)}
\,
e^{\mu+\sigma^2/2}.
\label{expected_lognormal}
\end{equation}
The value obtained from this formula for $u=10^6$ is in total agreement with the results of the simulations.
Subtracting $u$ to the formula we obtain $\epsilon_{ln}(u)$,
which is represented in Fig. \ref{Fig5}(a) as a function of $u$.

One could be tempted to reduce the resulting value of $\langle x | x \ge 10^6\rangle$
by introducing a much faster decay for very high values of $x$ (e.g., at $x=h$).
However, Ref. \cite{Corral_comment_CT2} shows that the results depend, obviously,
not only on the value of $h$ but also on the form of this fast decay,
and the decay cannot be postulated ad-hoc.
Note also that the estimation of $\langle x | x \ge u\rangle_{ln}$
from empirical or from simulated sampled is, naturally, a random variable
(for fixed $u$).
Figure \ref{Fig5}(a) illustrates how  
the median of $\langle x | x \ge u\rangle_{ln}$
(around where most simulations gather)
is close to the empirical result,
but the previous calculation shows that 
the mean is far, and much higher.

%DECIR ALGO MAS???

The results in this subsection 
highlight the importance of the underlying statistical model,
as the results may depend more on the assumptions contained in the model
than on the empirical data.
The estimations are crude because are based on crude data, 
just a one-dimensional random variable.
Knowledge of the dynamics of the growth of the death toll with time
until its final value $x$
(on a daily, or monthly basis, etc.) for the historical data (in other words, 
knowledge of the ``avalanche profile'')
would provide more valuable information to improve the current estimation.
Other limitations are explained in the final section.

\section{Discussion on the possibility of a fat tail in epidemic fatalities}

%We have provided enough evidence that 
%Our results with the 

The fact that the truncated log-normal distribution fits well the epidemic data 
is not a unique attribute of this distribution and probably 
other theoretical distributions, fat-tailed or not,
can do a similar good job;
that is, there are candidate distributions, 
such as the stretched exponential or the Weibull,
that could reproduce the empirical results well enough.
This means that
the ``true'' probability distribution describing the number of fatalities of epidemics
cannot be established from a purely statistical analysis.

Fortunately, 
physical insights can shed light on this problem.
Assuming 
a very simple (mean-field) model in which infections propagate following a Galton-Watson
stochastic branching process \cite{Harris,branching_biology,Corral_FontClos}, 
with a number of fatalities that is a fixed fraction of the number of infections
(constant, deterministic infection fatality risk \cite{Hill_epidemics}),
%OJO A ESTO, NO SERA UN RATE!!! ???
and
identifying the branching ratio with the basic reproductive number, $R_0$, 
it is immediate to see that
$R_0 < 1$ leads to rather small epidemics
(few number of fatalities, given by an exponential tail for $x$ ), 
whereas for $R_0 > 1$ two scenarios are possible
starting from a single individual:
again, few fatalities (``good-luck'' case), 
or an infinite number of fatalities (``bad-luck'', in an infinite system).
It is only at the critical point, $R_0=1$, where the number of fatalities is fat tailed,
with an exponent $\alpha=1/2$ (and this is known at least since the 1940s \cite{Harris},
provided that the distribution of contagions arising directly from one individual
has a finite second moment).
% Otter 1949

For a sequence of historical epidemics, 
the distribution of the resulting number of fatalities will be a mixture of
subcritical ($R_0<1$), critical, and supercritical ($R_0>1$) distributions, 
weighted by the distribution of $R_0$, whose density is denoted here by $\rho(R_0)$; thus, 
$$
f(x) = \int  f(x|R_0)\rho(R_0) dR_0.
$$
For large $x$, only the critical and supercritical regimes need to be taken into account
(as in subcriticality large values of $x$ are totally negligible). 
Moreover, in a supercritical situation, whenever $x$ reaches large enough values, 
one expects that social interventions to fight the epidemic are implemented;
if these are effective the value of $R_0$ should decrease, 
slightly below one in the ideal case
(and we should deal instead with $R_T$, with $T$ the internal time of the epidemic).
In this way, the contention of the epidemic triggers a feedback mechanism that,
when $x$ is large,
sets the value of $R_T$ close to the critical point of the model.
The situation is that of self-organized critical phenomena \cite{Bak_book,Zapperi_branching,Watkins_25years}.
Under these circumstances, one would expect a power-law tail for large $x$, 
with an exponent $\alpha=1/2$ \cite{Corral_FontClos}.

This shows in a simple scenario how a power-law tail is feasible,
in agreement with Cirillo and Taleb \cite{Cirillo_Taleb}.
Nevertheless, the situation just described is highly idealized, and the precise results would depend on the ``natural'' distribution of $R_0$ for contagious diseases and the dynamics of $R_T$ under mitigation measures.
Finite-size effects should also be taken into account properly 
\cite{GarciaMillan,Corral_garcia_moloney_font}.

In addition, needless to say, the Galton-Watson model is too simplistic, and other models, beyond mean field,
may lead at least to a different value of the exponent $\alpha$.
For example, superspreading phenomena for which the number of contagions triggered directly by a single individual were power-law distributed \cite{Wong_superspreading}
would lead, counterintuitively, to larger values of $\alpha$ 
($\alpha>1/2$, still fat tails, but thinner than in the Galton-Watson model \cite{Saichev_Sornette_branching}). 
In short, 
%we have seen that, 
from a theoretical point of view, 
it seems reasonable that the epidemic size distribution is fat tailed,
but, in any case, 
the hypothetical theoretical support
does not make the supposed empirical 
evidence provided by Ref. \cite{Cirillo_Taleb} more valid.

%The influence of finite size effects has been discussed!!!

%DECIR QUIEN FUE EL PRIMERO QUE ENCONTRO EL EXPONENTE 3/2... !!!!!! Otter?

%C y T dicen que la distribucion seria Gaussiana???
%No!! Solo dicen que los accidentes de coche lo son!
%Pero el editor de Nat Phys lo entendio asi!!!

\section{Conclusion}

We have shown that there is not enough empirical evidence %that supports 
%that the distribution of fatalities in epidemics along history is fat-tailed.
that the fatalities caused by epidemics along history follow
a fat-tailed distribution.
A log-normal tail (lacking fat-tailness) 
is able to replicate all sort of metrics used before \cite{Cirillo_Taleb} to support
fat-tailness in the empirical record. % of fatalities in epidemics.
For sure, some other not fat-tailed distributions could fit the data
similarly well as the log-normal.
What our work shows is the importance of considering alternative probability models
when fitting heavy-tailed distributed data (which is different from fat-tailed data \cite{Voitalov_krioukov}), 
as well as the key role of computer simulations to contrast the validity 
of theoretical results when the number of data is not infinite.

Summarizing, log-normal tailed simulated data, 
in the same way as the empirical data,
has:
a mean excess size $\epsilon(u)$
that increases with a lower cut-off in size $u$;
a maximum-to-sum ratio that does not tend to zero as the number of events increases
(up to $N=72$);
and
a power-law tail exponent $\alpha$ that also increases with $u$
(and which can be erroneously associated to a positive tail index $\xi$, 
despite the fact the log-normal belongs to the 
Gumbel maximum domain of attraction and should have $\xi =0$ \cite{Coles};
obviously, 
the calculation of $\xi$ from the maximum-likelihood (Hill) 
%power-law exponent 
estimator
implicitly assumes $\xi >0$).
The agreement between empirical data and the log-normal simulations is not only qualitative but quantitative for the three metrics,
although there is substantial scattering in the outcome
of the simulation results (due to the small value of $N$).

%%EXPLICAR HILL???

Moreover, the uniformly most powerful unbiased test based on the logarithmic coefficient of variation
\cite{Malevergne_Sornette_umpu}
shows that the power-law fit is preferred for the top 21 events of the empirical data, 
but this preference for a power-law tail is also shown for log-normal synthetic events;
this is due to the well-known fact that the power law can be considered as a special case of a log-normal
tail \cite{Malevergne_Sornette_umpu}. 
%So, and this is something that it is not so well known, the log-normal distribution, 
%belonging to the Gumbel maximum domain of attraction, has a special power-law limit, 
%belonging to the Fr\'echet maximum domain of attraction.
%In other words, when it is commonly stated that ``log-normal is Gumbel''
%the special power-law limit of the truncated log-normal 
%($\sigma^2 \rightarrow \infty$ and $\mu \rightarrow -\infty$)
%is implicitly disregarded.

Our results also demonstrate that the risks brought by log-normally tailed phenomena
can be enormous.
Still, 
one may argue that from the point of risk management it is more conservative
to take the power law (which gives a larger probability for the most extreme events)
than the log-normal tail.
This is true, but constitutes a different problem, which could be addressed even without 
any statistical modelling.
For example,
using the empirical data of Ref. \cite{Cirillo_Taleb}, 
if an epidemic reaches 1000 fatalities 
then it has a non-negligible probability ($1/72=0.014$) of yielding 138 million fatalities.
After reaching 10,000 fatalities
this probability further increases (to $(1/72)/(55/72)\simeq 0.02$), 
and so on
(of course, in this context extrapolation would not be possible and the probability of having an event with more fatalities than the Black Death cannot be computed).
As the empirical data are rather incomplete, 
the previous numbers should not be considered truly reliable, 
but the same happens with the conclusions derived from any 
statistical model fitted to those data.

In fact,
an important limitation when studying the distribution of fatalities in epidemics 
comes from the available data, 
not only because of the small sample size ($N=72$ in the data of Ref. \cite{Cirillo_Taleb}),
%% used here),
but also from the incompleteness of the data (with a bias in favor of very large events
that resampling \cite{Cirillo_Taleb} cannot correct)
and from the lack of homogeneity in time (with just one event, the Black Death, between 750 and 1450,
and 11 events since 2008 in the data of Ref. \cite{Cirillo_Taleb}).

Notice that the data are inhomogeneous in an additional way: 
epidemics in the Middle Ages and in the 21st century are not comparable in the sense that 
they propagate differently and that the measures implemented for their contention 
should be more effective nowadays.
If one mixes historical epidemic data with contemporary data
what one obtains is, obviously, a mixture of distributions.
This is not wrong {\it per se}, but one needs to have in mind
for which reason one needs such knowledge
in order to interpret the results properly.
At the end, statistics derived from epidemic data of previous times
have limited applicability nowadays 
(except if we faced pandemics with the same errors than in the Middle Ages).

In any case, one
can dig a little in the existing records and find many more historical events.
As an example, Villalba \cite{Villalba} reported several epidemics in Spain with more than 10,000 fatalities
that are not considered in the data of Ref. \cite{Cirillo_Taleb}
% https://twitter.com/_MiguelHernan/status/1263239450240393222
% https://books.google.es/books?id=EvsBb03_d_UC&printsec=frontcover&hl=es#v=onepage&q&f=false
%
%con mas de 10000 muertos
%
(these missing Spanish epidemics took place in 
1283,
% 1333, %peste negra
% 1348, %peste negra
1394,
1490, % Sweating sickness?? NO, no llego a España 1489?? Tifus? Esta en wikipedia pero no en C&T
1564, % London plague?? NO, solo Londres
1589,
%1596, Plague in Spain
1637,
%1647, % Great plague of Seville
%1649, % id
1726,
1741,
1784, and
1800).
For sure, 
there is nothing special about Spain,
and
other countries can contribute more or less in the same
way with more ``hidden'' epidemics.
% searching "mil"
Nevertheless, the compilation of a reliable record for historical epidemics is something
that should not be done by probabilists, statisticians, or physicists,
and needs to be carefully undertaken by true epidemiologists and historians.
We urge here for the necessity of such an important endeavor.

%OJO, LO DE LOS 80 MILLONES ESTA BIEN? NO CUADRA CON LA FIG 3A!!!!!
%OJO, EL PROBLEMA ES QUE HAY MUCHISIMA DISPERSION, INCLUSO PARA LA LOGNORMAL

%NO HACE FALTA HACER ESTADISTICA, COGER DATOS EMPIRICOS 1/72!!!! 

%% 800,000 en 4 epidemias de cholera en el 19th century
%% https://es.wikipedia.org/wiki/Epidemias_de_c%C3%B3lera_en_Espa%C3%B1a

%
%\section{Conclusions}
%
%We have shown how the probability distribution of the number of fatalities of historical epidemics can be well explained by a log-normal distribution, 
%which is a distribution that is empirically similar to the power law
%but quite different from a theoretical point of view (in particular, 
%the mean and all moments of a log-normal distribution are well defined).
%%IMPORTANCE OF SIMULATIONS!!

%ANYADIR CITA AL OTRO PAPER DE EDUARD, CON SALJE Y PLANES!!!
%\cite{Salje_Vives}

\section{Acknowledgments}

I acknowledge 
Isabel Serra for discussions
and
Miguel Hern\'an and Diego Ramiro Fari\~nas for drawing my attention to Ref. \cite{Villalba}.
And also 
support from projects
FIS2015-71851-P and
PGC-FIS2018-099629-B-I00
%Red de Excelencia MAT2015-69777-REDT, 
%and 
%Mar\'{\i}a de Maeztu Program 
%%for Units of Excellence in R\&D 
%MDM-2014-0445,
from Spanish MINECO and MICINN.
I regret that it has been not possible to discuss these results
with my colleague P. Puig, due to his problems with COVID-19.

\section{Appendix I: Moments and logarithmic coefficient of variation of
the truncated log-normal distribution}

The logarithmic coefficient of variation of a random variable $x$ above
a threshold $u$ is defined as the standard deviation of 
$\ln(x/u)$ divided by the expected value of $\ln(x/u)$, i.e., 
$$
C=\frac{\sqrt{\langle \ln^2 (x/u)\rangle - \langle \ln (x/u)\rangle^2}}{\langle \ln (x/u)\rangle},
$$
where it is implicit that $x\ge u$.
For the case of a truncated log-normal distribution with parameters $\mu$ and  
$\sigma^2$, given by Eq. (\ref{Eqlognormal}), the following change of variables
$t=(\ln x - \mu)/\sigma$ leads to 
$$
C=\frac{\sigma\sqrt{\langle t^2 \rangle-\langle t \rangle^2}}
{\mu-\ln u+\sigma \langle t \rangle},
$$ 
and $t$ turns out to follow a ``tipified''
truncated normal (tn) distribution 
with density
$$
f_{tn}(t)=\frac{e^{-t^2/2}} Z,
\mbox{ with }
Z= \sqrt{2\pi } \times \frac 1 2\mbox{erfc}\left(\frac{\ln u -\mu}{\sqrt{2} \sigma}\right),
$$
for $t\ge (\ln u -\mu)/\sigma$.
We call the distribution of $t$ tipified because the parameters $\mu$ and $\sigma$
have been transformed to take values zero and one, 
but these values are not the mean and variance of $t$.
Direct calculation of the moments of $t$ is straightforward, leading to
$$
\left \langle \ln \frac x u \right\rangle = \tilde \mu+\sigma \langle t\rangle
=\tilde \mu +\sigma \frac{e^{-(\tilde\mu/\sigma)^2/2}}Z
$$
with $\tilde\mu =\mu-\ln u$.
This is the same as Eq. (\ref{xilognormal}) for $\langle \alpha^{-1}\rangle_{ln}$. Also, 
$$
\langle \ln^2 (x/u)\rangle - \langle \ln (x/u)\rangle^2 =
\sigma^2(\langle t^2 \rangle - \langle t\rangle^2)
=
\sigma^2 \left(
1 - \frac{\tilde \mu} \sigma \frac{e^{-(\tilde\mu/\sigma)^2/2}}Z
-\frac{e^{-(\tilde\mu/\sigma)^2}}{Z^2}
 \right)
$$
and therefore, the logarithmic coefficient of variation is given
(for the truncated log-normal distribution) by
\begin{equation}
C=\frac{\sigma\sqrt{
1 - {\tilde \mu} \sigma^{-1} {e^{-(\tilde\mu/\sigma)^2/2}}/Z
-{e^{-(\tilde\mu/\sigma)^2}}/{Z^2}
}}
{\tilde \mu +\sigma {e^{-(\tilde\mu/\sigma)^2/2}}/Z
}.
\label{finalC}
\end{equation}
As $\mu$ and $\sigma$ are fixed (determined from the fit of the empirical data),
$C$ depends only on $\ln u$.
Figure \ref{Fig5}(c) represents $C$ as a function of the number of points in the tail, 
which, for each value of $u$, are estimated as $n=N S_{ln}(u)$, with $N=72$ and 
\begin{equation}
S_{ln}(u)= \frac{
 \mbox{erfc}\left(\frac{\ln u -\mu}{\sqrt{2} \sigma}\right)
}{
 \mbox{erfc}\left(\frac{\ln 10^3 -\mu}{\sqrt{2} \sigma}\right)
}.
\label{finalC2}
\end{equation}

\newpage
\section{Appendix II: Logarithmic coefficient of variation computation in R language}

Although the codes used in this research have been developed in FORTRAN 77,
we present a simple alternative in R for the logarithmic coefficient of variation.
This can be used as a double check of our results.
The program below simulates a truncated log-normal sample, 
draws the histogram using logarithmic binning
(corresponding to Fig. \ref{Fig2}(a))
and draws the logarithmic CV plot 
(corresponding to Fig. \ref{Fig5}(c)).
The (self-contained) R code follows
\begin{verbatim}
mu<-10.43; sigma<-3.60
N<-72; x[1:N]<-0
for (i in 1:N){
    while (x[i]<=1000) {
        x[i]<-exp(rnorm(1,mean=mu,sd=sigma)) }}

histog_log<-hist(log(x),probability = 'T',col='blue')
plot(exp(histog_log$mids),histog_log$density/exp(histog_log$mids),log='xy',type='p')

install.packages('ercv')
require(ercv)
cvplot(log(x),conf.level=0.85)
\end{verbatim}
 
The R package ercv, used to draw to log CV plot, 
was published in Ref. \cite{Morina_R}.

\newpage

%\bibliographystyle{unsrt}   % por orden de cita
%\bibliography{C:/Users/acorr/Dropbox/p1_lemmas/biblio}

\begin{thebibliography}{10}

\bibitem{Bak_book}
P.~Bak.
\newblock {\em How Nature Works: The Science of Self-Organized Criticality}.
\newblock Copernicus, New York, 1996.

\bibitem{Mitz}
M.~Mitzenmacher.
\newblock A brief history of generative models for power law and lognormal
  distributions.
\newblock {\em Internet Math.}, 1 (2):226--251, 2004.

\bibitem{Newman2004a}
M.~E.~J. Newman.
\newblock Power laws, {Pareto} distributions and {Zipf's} law.
\newblock {\em Contemporary Physics}, 46:323--351, 2005.

\bibitem{Sornette_critical_book}
D.~Sornette.
\newblock {\em Critical Phenomena in Natural Sciences}.
\newblock Springer, Berlin, 2nd edition, 2004.

\bibitem{Thurner_book}
S.~Thurner, R.~Hanel, and P.~Klimek.
\newblock {\em Introduction to the Theory of Complex Systems}.
\newblock Oxford University Press, New Delhi, 2018.

\bibitem{Cirillo_Taleb}
P.~Cirillo and N.~N. Taleb.
\newblock Tail risk of contagious diseases.
\newblock {\em Nature Phys.}, 16:606--613, 2020.

\bibitem{White}
E.~P. White, B.~J. Enquist, and J.~L. Green.
\newblock On estimating the exponent of power-law frequency distributions.
\newblock {\em Ecol.}, 89:905--912, 2008.

\bibitem{Bauke}
H.~Bauke.
\newblock Parameter estimation for power-law distributions by maximum
  likelihood methods.
\newblock {\em Eur. Phys. J. B}, 58:167--173, 2007.

\bibitem{Clauset}
A.~Clauset, C.~R. Shalizi, and M.~E.~J. Newman.
\newblock Power-law distributions in empirical data.
\newblock {\em SIAM Rev.}, 51:661--703, 2009.

\bibitem{corral_nuclear}
A.~Corral, F.~Font, and J.~Camacho.
\newblock Non-characteristic half-lives in radioactive decay.
\newblock {\em Phys. Rev. E}, 83:066103, 2011.

\bibitem{Corral_Deluca}
A.~Deluca and A.~Corral.
\newblock Fitting and goodness-of-fit test of non-truncated and truncated
  power-law distributions.
\newblock {\em Acta Geophys.}, 61:1351--1394, 2013.

\bibitem{Hanel_power_laws}
Hanel R., Corominas-Murtra B., Liu B., and Thurner S.
\newblock Fitting power-laws in empirical data with estimators that work for
  all exponents.
\newblock {\em PLoS ONE}, 12(2):e0170920, 2017.

\bibitem{Corral_Gonzalez}
A.~Corral and A.~Gonz\'alez.
\newblock Power law distributions in geoscience revisited.
\newblock {\em Earth Space Sci.}, 6(5):673--697, 2019.

\bibitem{Voitalov_krioukov}
I.~{Voitalov}, P.~{van der Hoorn}, R.~{van der Hofstad}, and D.~{Krioukov}.
\newblock Scale-free networks well done.
\newblock {\em Phys. Rev. Research}, 1:033034, 2019.

\bibitem{Corral_comment_cirillo_taleb}
A.~Corral.
\newblock Scientific comment on ``{Tail} risk of contagious diseases''.
\newblock {\em arXiv}, 2007.06876, 2020.

\bibitem{Wikipedia_epidemics}
Wikipedia.
\newblock List of epidemics.
\newblock {\em https://en.wikipedia.org/wiki/List$\_$of$\_$epidemics}.

\bibitem{list_epidemics}
ListFist.
\newblock List of epidemics compared to coronavirus.
\newblock {\em
  https://listfist.com/list-of-epidemics-compared-to-coronavirus-covid-19}.

\bibitem{Corral_garcia_moloney_font}
A.~Corral, R.~Garcia-Millan, N.~R. Moloney, and F.~Font-Clos.
\newblock Phase transition, scaling of moments, and order-parameter
  distributions in {Brownian} particles and branching processes with
  finite-size effects.
\newblock {\em Phys. Rev. E}, 97:062156, 2018.

\bibitem{Corral_Paczuski}
A.~Corral and M.~Paczuski.
\newblock Avalanche merging and continuous flow in a sandpile model.
\newblock {\em Phys. Rev. Lett.}, 83:575--578, 1999.

\bibitem{footnote_survival}
Note that the name surival or survivor function can be confusing. {It} makes a
  lot of sense when the random variable is a failure time or a lifetime, but
  not when we are dealing with other variables, as fatalities. {Then}, the
  interpretation of the survival function here cannot be the same as in
  reliability theory and survival analysis.

\bibitem{Abramowitz}
M.~Abramowitz and I.~A. Stegun, editors.
\newblock {\em Handbook of Mathematical Functions}.
\newblock Dover, New York, 1965.

\bibitem{Malevergne_Sornette_umpu}
Y.~Malevergne, V.~Pisarenko, and D.~Sornette.
\newblock Testing the {Pareto} against the lognormal distributions with the
  uniformly most powerful unbiased test applied to the distribution of cities.
\newblock {\em Phys. Rev. E}, 83:036111, 2011.

\bibitem{Corral_Arcaute}
A.~Corral, F.~Udina, and E.~Arcaute.
\newblock Truncated lognormal distributions and scaling in the size of
  naturally defined population clusters.
\newblock {\em Phys. Rev. E}, 101:042312, 2020.

\bibitem{Corral_comment_CT2}
A.~Corral.
\newblock Finite-size scaling versus dual random variables and shadow moments
  in the size distribution of epidemics.
\newblock {\em arXiv}, 2011.04316, 2020.

\bibitem{Kalbfleisch2}
J.~D. Kalbfleisch and R.~L. Prentice.
\newblock {\em The Statistical Analysis of Failure Time Data}.
\newblock Wiley, Hoboken, NJ, 2nd edition, 2002.

\bibitem{Schroeder}
M.~Schroeder.
\newblock {\em Fractals, Chaos, Power Laws}.
\newblock Freeman, New York, 1991.

\bibitem{Bouchaud_Georges}
J.-P. Bouchaud and A.~Georges.
\newblock Anomalous diffusion in disordered media: statistical mechanisms,
  models and physical applications.
\newblock {\em Phys. Rep.}, 195:127--293, 1990.

\bibitem{Corral_csf}
A.~Corral.
\newblock Scaling in the timing of extreme events.
\newblock {\em Chaos. Solit. Fract.}, 74:99--112, 2015.

\bibitem{Castillo}
J.~{del Castillo} and P.~Puig.
\newblock The best test of exponentiality against singly truncated normal
  alternatives.
\newblock {\em J. Am. Stat. Assoc.}, 94:529--532, 1999.

\bibitem{pawitan2001}
Y.~Pawitan.
\newblock {\em In All Likelihood: Statistical Modelling and Inference Using
  Likelihood}.
\newblock Oxford UP, Oxford, 2001.

\bibitem{Baro_Vives}
J.~Bar\'o and E.~Vives.
\newblock Analysis of power-law exponents by maximum-likelihood maps.
\newblock {\em Phys. Rev. E}, 85:066121, 2012.

\bibitem{Coles}
S.~Coles.
\newblock {\em An Introduction to Statistical Modeling of Extreme Values}.
\newblock Springer, London, 2001.

\bibitem{Salje_Vives}
E.~K.~H. Salje, A.~Planes, and E.~Vives.
\newblock Analysis of crackling noise using the maximum-likelihood method:
  Power-law mixing and exponential damping.
\newblock {\em Phys. Rev. E}, 96:042122, 2017.

\bibitem{Harris}
T.~E. Harris.
\newblock {\em The Theory of Branching Processes}.
\newblock Dover, New York, 1989.

\bibitem{branching_biology}
M.~Kimmel and D.~E. Axelrod.
\newblock {\em Branching Processes in Biology}.
\newblock Springer-Verlag, New York, 2002.

\bibitem{Corral_FontClos}
A.~Corral and F.~Font-Clos.
\newblock Criticality and self-organization in branching processes: application
  to natural hazards.
\newblock In M.~Aschwanden, editor, {\em Self-Organized Criticality Systems},
  pages 183--228. Open Academic Press, Berlin, 2013.

\bibitem{Hill_epidemics}
A.~L. Hill.
\newblock The math behind epidemics.
\newblock {\em Phys. Today}, 73(11):28--34, 2020.

\bibitem{Zapperi_branching}
S.~Zapperi, K.~B. Lauritsen, and H.~E. Stanley.
\newblock Self-organized branching processes: Mean-field theory for avalanches.
\newblock {\em Phys. Rev. Lett.}, 75:4071--4074, 1995.

\bibitem{Watkins_25years}
N.~W. Watkins, G.~Pruessner, S.~C. Chapman, N.~B. Crosby, and H.~J. Jensen.
\newblock 25 years of self-organized criticality: Concepts and controversies.
\newblock {\em Space Sci. Rev.}, 198:3--44, 2016.

\bibitem{GarciaMillan}
R.~Garcia-Millan, F.~Font-Clos, and A.~Corral.
\newblock Finite-size scaling of survival probability in branching processes.
\newblock {\em Phys. Rev. E}, 91:042122, 2015.

\bibitem{Wong_superspreading}
F.~Wong and J.~J. Collins.
\newblock Evidence that coronavirus superspreading is fat-tailed.
\newblock {\em Proc. Natl. Acad. Sci. USA}, 117(47):29416--29418, 2020.

\bibitem{Saichev_Sornette_branching}
A.~Saichev, A.~Helmstetter, and D.~Sornette.
\newblock Power-law distributions of offspring and generation numbers in
  branching models of earthquake triggering.
\newblock {\em Pure Appl. Geophys.}, 162:1113--1134, 2005.

\bibitem{Villalba}
J.~{de Villalba}.
\newblock {\em Epidemiolog{\'\i}a espa{\~n}ola o Historia cronol{\'o}gica de
  las pestes, contagios, epidemias y epizootias que han acaecido en Espa{\~n}a
  desde la venida de las cartagineses hasta el a{\~n}o 1801}.
\newblock Imprenta de Ferm{\'\i}n Villalpando, Madrid, 1803.

\bibitem{Morina_R}
J.~{del Castillo}, I.~Serra, M.~Padilla, and D.~Mori{\~n}a.
\newblock {Fitting Tails by the Empirical Residual Coefficient of Variation:
  The ercv Package}.
\newblock {\em {R Journal}}, 11(2):56--68, 2019.

\end{thebibliography}

%
%\section{MORE}
%
%SE PODRIA COMPARAR NO CON PARETO, SINO CON UNA COLA POWER LAW CON BODY BOOTSTRAPEDADO DE EMPIRICAL FINDINGS.!!
%
%EN PARTICULAR MIRAR LA ESTADISTICA DEL MAXIMO!!!!
%(ES BLOCK-MAXIMA PERO CON BLOQUES DE TAMANYO 1. NO IMPORTA, NO SE APLICA LA TEORIA ASINTOTICA, PERO TANTO DA!!)
%
%
%
%%DECIR QUE LAS MEDIDAS DE CONTENCION SON DIFERENTES AHORA QUE PARA LA PESTE NEGRA!
%
%LOS MOMENTOS FANTASMAS DE TALEB SE PUEDEN CALCULAR FACILMENTE POR SIMULACION!!!
%
%%ANYADIR UNA DISCUSION SOBRE POR QUE TENDRIA QUE SER UNA POWER LAW: 
%%VALORES DE R0, NO SABEMOS SU DISTRIBUCION, CUANDO R0 ES MAYOR QUE UNO HAY UN ESFUERZO POR BAJARLO HASTA 1, 
%%PERO ESO ES SUFICIENTE PARA DAR UNA POWER LAW...???
%
%%NO SOLO DEPENDE DE R0, TAMBIEN DEL NUMERO DE SEMILLAS, PRIMERA GENERACION, PACIENTE0...
%
%\section{Conclusions}
%
%%IMPORTANCE OF SIMULATIONS!!
%
%ANYADIR CITA AL OTRO PAPER DE EDUARD, CON SALJE Y PLANES!!!
%\cite{Salje_Vives}
%
%citar tambien Dutta, K. Perry, J. (2006). A Tale of Tails: An Empirical Analysis of Loss
%Distribution Models for Estimating Operational Risk Capital. Federal Reserve
%Bank of Boston. Working Paper 06-13.
%
%% HACER APENDICE MACHACANDO A TALEB. Esto sera otro paper ya!!
%
%
%%CITAR CITA DE HERNAN DE 1800 Y PICO??? Y DIEGO RAMIRO???
%%MEJOR EN LAS CONCLUSIONES, URGIENDO A CREAR UN BUEN DATA SET.
%
%
%
%%slow varying FUNCTION!!!! MIRAR DE PINTARLA Y VER QUE NO ES TAN SLOWLY VARYING!!
%%(OJO, PARA LA ACUMULADA??). No es conclusico, obviamente los ultimos 3 o 4 puntos salen "alineados"

\end{document}